\documentclass[sigconf,screen,nonacm]{acmart}
\usepackage{graphicx}
\usepackage{textcomp}
\usepackage{xcolor}
\usepackage{fancyhdr}
\usepackage{multirow}
\usepackage{comment}
\usepackage[normalem]{ulem}
\usepackage{algorithm2e}
\usepackage{xspace}
\usepackage{microtype}

\newcommand{\facebook}{Meta\xspace}

\newcommand{\CarbonExplorer}{Carbon Explorer\xspace}
\newcommand{\carbon}{CO\textsubscript{2} }

\setcopyright{acmlicensed}
\acmPrice{15.00}
\acmDOI{10.1145/3575693.3575754}
\acmYear{2023}
\copyrightyear{2023}
\acmSubmissionID{asplosb23main-p249-p}
\acmISBN{978-1-4503-9916-6/23/03}
\acmConference[ASPLOS '23]{Proceedings of the 28th ACM International Conference on Architectural Support for Programming Languages and Operating Systems, Volume 2}{March 25--29, 2023}{Vancouver, BC, Canada}
\acmBooktitle{Proceedings of the 28th ACM International Conference on Architectural Support for Programming Languages and Operating Systems, Volume 2 (ASPLOS '23), March 25--29, 2023, Vancouver, BC, Canada}
\received{2022-07-07}
\received[accepted]{2022-09-22}

\begin{document}
%%%%%%%%%%%---SETME-----%%%%%%%%%%%%%
\title[Carbon Explorer: A Holistic Framework for Designing Carbon Aware Datacenters]{Carbon Explorer: A Holistic Framework \\ for Designing Carbon Aware Datacenters}

\author{Bilge Acun}
\email{acun@meta.com}
\affiliation{%
  \institution{Meta}
  \country{USA}
}

\author{Benjamin Lee}
\email{leebcc@seas.upenn.edu}
\affiliation{%
  \institution{University of Pennsylvania, Meta}
  \country{USA}
}

\author{Fiodar Kazhamiaka}
\email{fiodar@stanford.edu}
\affiliation{%
  \institution{Stanford University}
  \country{USA}
}

\author{Kiwan Maeng}
\email{kwmaeng@meta.com}
\affiliation{%
  \institution{Meta}
  \country{USA}
}
\author{Udit Gupta}
\email{uditg@meta.com}
\affiliation{%
  \institution{Harvard University, Meta}
  \country{USA}
}
\author{Manoj Chakkaravarthy}
\email{mchakkar@meta.com}
\affiliation{%
  \institution{Meta}
  \country{USA}
}
\author{David Brooks}
\email{dbrooks@eecs.harvard.edu}
\affiliation{%
  \institution{Harvard University, Meta}
  \country{USA}
}
\author{Carole-Jean Wu}
\email{carolejeanwu@meta.com}
\affiliation{%
  \institution{Meta}
  \country{USA}
}

%%%%%%%%%%%%%%%%%%%%%%%%%%%%%%%%%%%%
\renewcommand{\shortauthors}{B. Acun, B. Lee, F. Kazhamiaka, K. Maeng, U. Gupta, M. Chakkaravarthy, D. Brooks, C.-J. Wu}

\begin{abstract}
Technology companies reduce their datacenters' carbon footprint by investing in renewable energy generation and receiving credits from power purchase agreements. Annually, datacenters offset their energy consumption with generation credits (Net Zero). But hourly, datacenters often consume carbon-intensive energy from the grid when carbon-free energy is scarce. Relying on intermittent renewable energy in every hour (24/7) requires a mix of renewable energy from complementary sources, energy storage, and workload scheduling. In this paper, we present the Carbon Explorer framework to analyze the solution space. We use Carbon Explorer to balance trade-offs between operational and embodied carbon, optimizing the mix of solutions for 24/7 carbon-free datacenter operation based on geographic location and workload. Carbon Explorer has been open-sourced at \url{https://github.com/facebookresearch/CarbonExplorer}. 
\footnote{Paper is accepted at \href{https://dl.acm.org/doi/10.1145/3575693.3575754}{ASPLOS'23.}}

\end{abstract}
\maketitle

%%%%%% -- PAPER CONTENT STARTS-- %%%%%%%%

\section{Introduction}

Carbon-free energy is essential for environmental sustainability. The UN's \textit{24/7 Carbon-Free Energy Compact} calls for an ambitious goal: "Every kilowatt-hour of electricity consumption is met with carbon-free electricity sources, every hour of every day, everywhere"~\cite{un_energy_compact}. In the US, the Department of Defense and the General Services Administration seek strategies for "supplying 24/7 carbon pollution-free electricity for the federal government"~\cite{dod_247}. Computing must do its part to address this societal challenge. Electricity consumed by datacenters is significant and growing rapidly. Datacenters world-wide consumed 205 TWh in 2018~\cite{Masanet:2020}, exceeding the annual consumption of countries such as Ireland and Denmark~\cite{owidenergy}. Information and communication technology may account for 7\% to 20\% of global electricity demand by 2030~\cite{jones2018stop, andrae2015global}.

\if 0
\begin{figure}[t]
\centering
%\includegraphics[width=\columnwidth]{img/renewables_historical_growth_2.png}
\includegraphics[width=1\columnwidth]{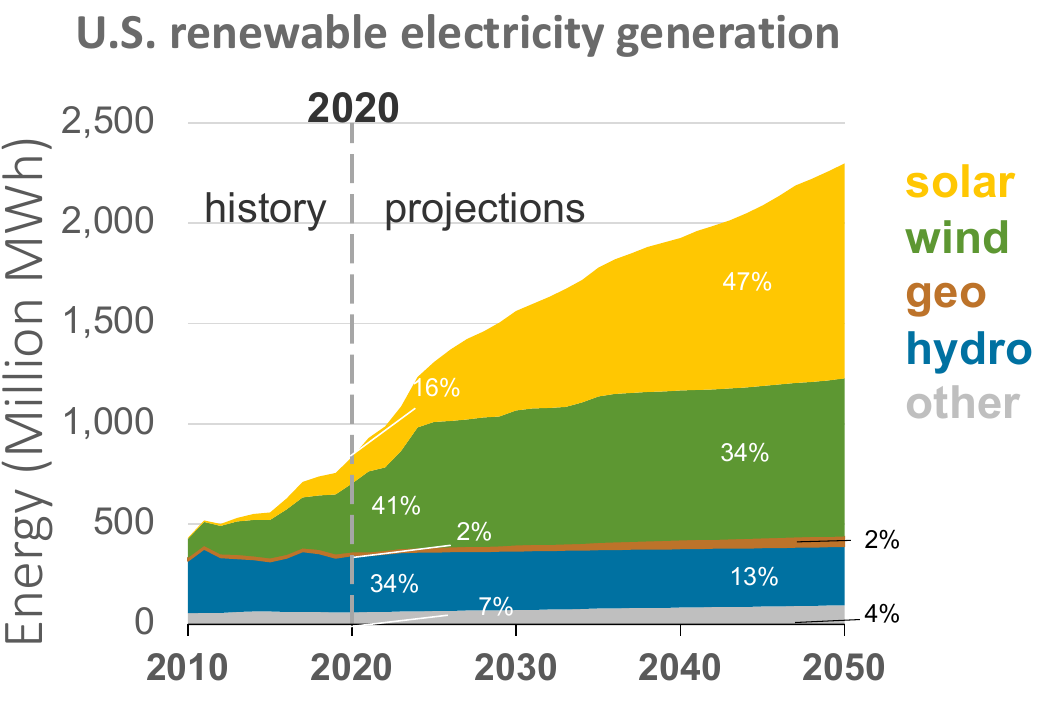}
\vspace{-0.3cm}
\caption{Historical and projection of growth of renewable energy in the US electricity grid [2010-2050].
81\% of the renewable energy will be comprised of solar and wind~\cite{eia_outlook}.
}
%\vspace{-0.1cm}
\label{motivation}
\end{figure}
\fi

\textbf{Net Zero versus 24~/~7.} 
At present, technology companies invest in renewable energy generation to offset datacenter energy consumption~\cite{facebook_sustainability, google_sustainability, microsoft_sustainability}. Amazon, Meta, and Google have collectively invested in 22 GW of renewable energy generation to meet {Net Zero} commitments. These investments align with broader efforts. In the United States, renewable energy generation is projected to increase from 20\% in 2020 to 42\% by 2050 as the nation pursues {Net Zero}~\cite{white_house_cop26}. Solar and wind comprise 47\% and 34\% of this renewable energy~\cite{eia_outlook}. 

\begin{figure}[t]
\centering
\includegraphics[width=0.75\columnwidth]{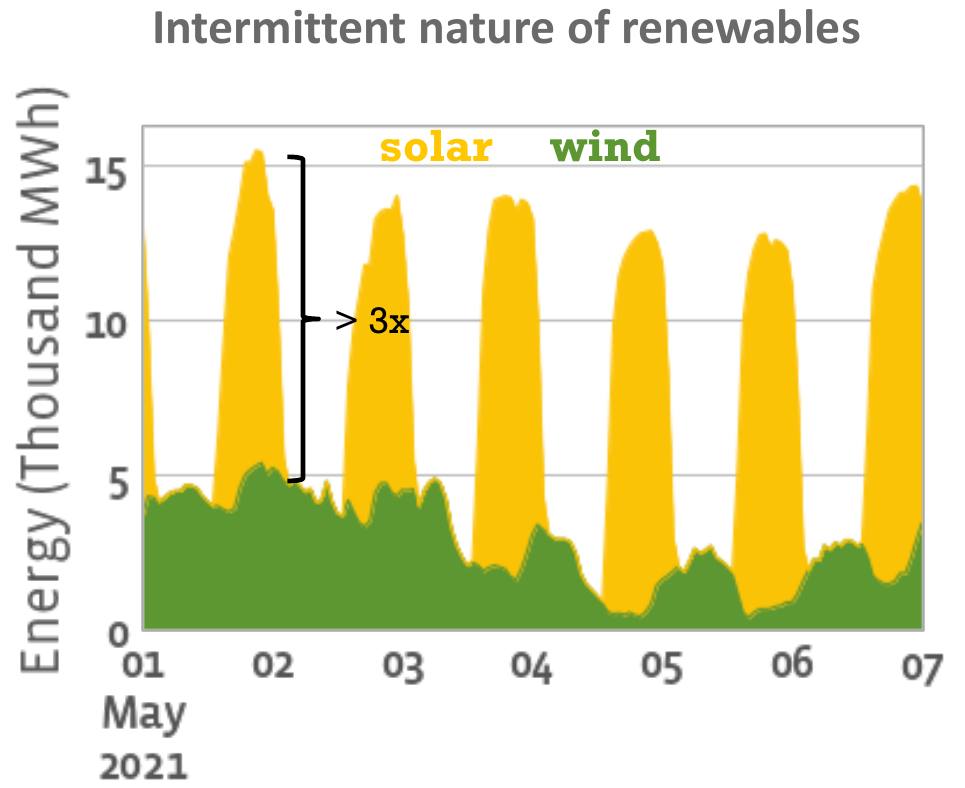}
\vspace{-0.1cm}
\caption{Hourly wind and solar energy generation in California grid during a week of time-frame.}
%\vspace{-0.1cm}
\label{motivation_fluctuations}
\end{figure}

Yet 24/7 carbon-free computing remains challenging because renewable energy supply fluctuates. The broad deployment of solar and wind farms will lead to increasingly severe hourly and seasonal supply fluctuations. Figure~\ref{motivation_fluctuations} highlights fluctuations when renewable generation comprises 33\% of the total. At times, the grid's supply of renewable energy may exceed demand, forcing inefficient curtailments that deactivate renewable energy generation in order to match supply with demand \cite{ciso_oversupply, chien20ees, chien18aims, lin20}. At other times, solar and wind energy is scarce. 

%and reduce congestion on the power transmission network

Under these conditions, datacenters consume carbon-intensive energy despite achieving Net Zero. Investments in wind and solar farms issue a renewable energy credit for every MWh generated. Annually, datacenters claim Net Zero by matching credits against energy consumed by their computation. Hourly, however, datacenters continue to emit carbon by consuming carbon-intensive energy (from gas or coal) when carbon-free supply (from wind or solar) is insufficient. Thus, more must be done to eliminate computing's hourly carbon emissions.

%Under these conditions, datacenters can claim Net Zero operations on an annual basis, matching total renewable energy credits generated by its wind and solar investments against the energy consumed by its computation, yet continue to generate carbon emissions on an hourly basis due to fluctuations in wind and solar supply. 

% Demand response
%
\textbf{Sustainability Design Space.}
24/7 carbon-free computing requires re-visiting classic questions in datacenter architecture and management. Architects have studied power infrastructure \cite{fan07} to maximize performance and utilization subject to power constraints \cite{fan16, govindan14, sakalkar20}. They have developed schedulers in which jobs request resources and schedulers allocate them \cite{hindman11,kubernetes}, perhaps with fairness and sharing incentives \cite{ghodsi11, zahedi14}. Architects must rethink these frameworks for carbon-free energy. 

Architects must first define parameterized solutions for coping  with intermittent renewable energy. And they require tools that reveal carbon-efficient combinations of these solutions. One solution is demand response, which schedules computation to align renewable energy supply and datacenter energy demand \cite{radovanovic2021carbon, wiesner2021let, bashir2021enabling, zhang21a, wierman14}. A second is energy storage, which reduces exposure to supply fluctuations \cite{liu21tsc, liu15}. Finally, further investment in wind and solar offers more carbon-free energy.

These solutions have implications for infrastructure. Datacenters will need additional servers that perform extra computation when carbon-free energy is abundant \cite{zhang21b}, batteries much larger than those in today's power infrastructure, and investments in energy generation that reflect the datacenter's location and relative availability of wind and solar. All of this infrastructure incurs embodied carbon costs, an important but under-explored aspect of datacenter sustainability. Architects must pursue coordinated design and management---spanning hardware architecture, datacenter systems, distributed software---to identify the carbon-optimal mix of solutions.

%This work shows that sustainability strategies require a cross-layer co-design (i.e. spaning across different layers from hardware, runtime and full datacenter system) and sustainability should not be approached in isolation for each datacenter layer.

\textbf{Carbon Explorer.} {Carbon Explorer} enables architects to optimize datacenters for 24/7 carbon-free computing (Section~\ref{2_carbon_explorer}). It consumes data about datacenter energy demands and renewable energy supplies across diverse geographic locations (Section~\ref{3_dc_demand_supply}). Moreover, it estimates the effect of investments in renewable energy using grid data, in energy storage using physically accurate battery models, and in server capacity needed for demand response scheduling using production datacenter traces (Section~\ref{4_carbon_aware_dc}). 

Carbon Explorer takes a holistic approach to datacenter design, balancing reductions in operational carbon against increases in embodied carbon incurred when manufacturing servers, batteries, and wind/solar farms. Although it seeks to maximize the number of hours using carbon-free energy (\textit{i.e.}, coverage), Carbon Explorer occasionally discovers that embodied carbon costs outweigh operational carbon reductions and identifies a solution that does not achieve 100\% coverage but is carbon-optimal. Specifically, we  report several key findings (Section~\ref{5_evaluation}). 

\begin{itemize}
    
    \item \textbf{Site Selection.} Among 13 regions where \facebook sites datacenters, Iowa and Nebraska (majorly wind) as well as hybrid regions (wind and solar) such as Texas are best for minimizing carbon costs because their supply valleys (days in the year with lowest levels of renewable energy supply) are shallowest.
    
    \item \textbf{Renewables Only.} Relying on renewable energy for coverage produces diminishing returns. Datacenters require 5$\times$ more renewables to increase coverage from 95\% to 99.9\% than from 0\% to 95\%. Accounting for embodied carbon in wind/solar farms, carbon-optimal designs achieve 37\% to 97\% coverage.
    
    \item \textbf{With Batteries.} Batteries permit datacenters to reach 100\% coverage for four regions and 99\% coverage for majority of the rest. Batteries must be large enough for a few hours of computation, but embodied carbon costs can be justified by reductions in operational carbon.
    
    \item \textbf{With Scheduling.} Demand response increases coverage by 1\%--22\% depending on the region. Scheduling requires 6\% to 76\% additional servers to support deferred computation, but their embodied carbon costs can be justified by reductions in operational carbon when 40\% of workloads are flexible.
    
    \item \textbf{All Together.} Demand response reduces battery capacity required and makes 100\% coverage optimal for five regions and above 99\% for rest of the regions except OR.
    
\end{itemize}

To further research in carbon-aware datacenter design and management, the Carbon Explorer framework has been made available at \url{https://github.com/facebookresearch/CarbonExplorer}.

%\textbf{Qualifications.} \

%Climate change is an existential crisis. We hope \CarbonExplorer will enable future works to deploy carbon-optimal technologies and achieve environmentally-sustainable computing in the years to come. 

\begin{figure*}[ht]
\centering
\includegraphics[width=2\columnwidth]{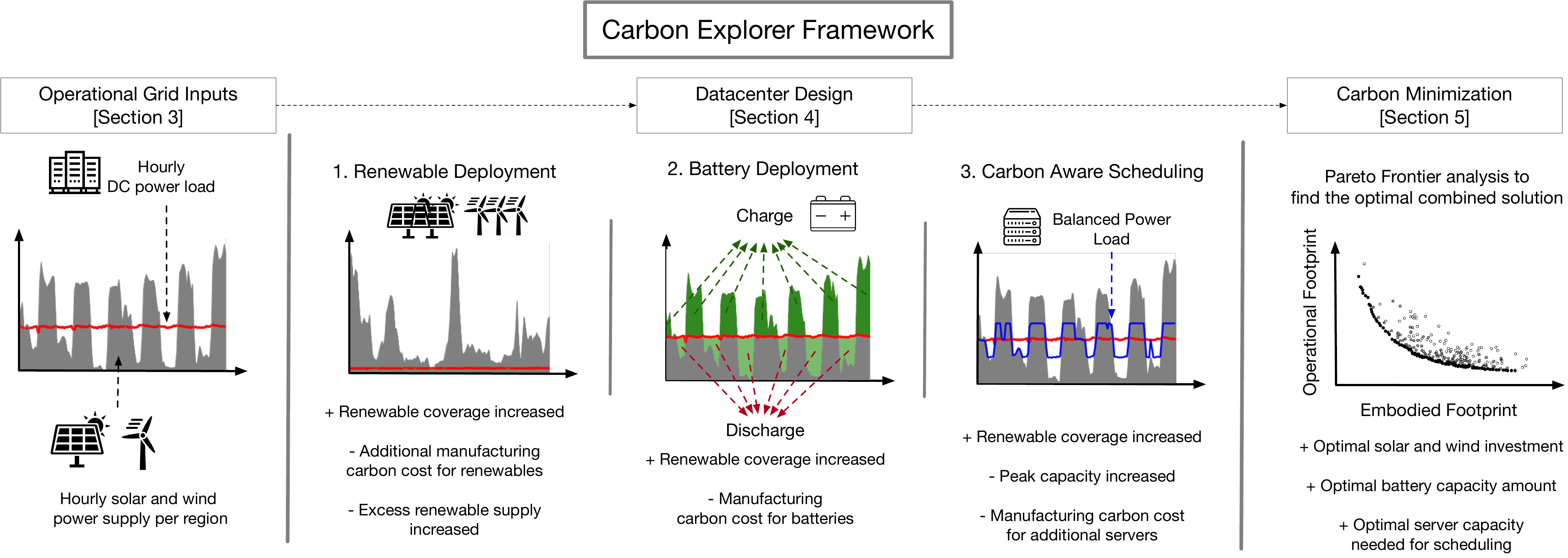}
\caption{Design Overview for \CarbonExplorer. \CarbonExplorer considers important characteristics, such as time-series power demand of large-scale datacenters and renewable energy availability on the power grids, as inputs. \CarbonExplorer characterizes the design space across renewable energy investments, energy storage, and computation shifting. \CarbonExplorer provides \textit{quantitative measures for strategies to achieve carbon-minimum settings}.}
\label{framework}
\end{figure*}

\section{Carbon Explorer}
\label{2_carbon_explorer}
Figure~\ref{framework} illustrates \textit{\CarbonExplorer}, a design space exploration framework that takes a holistic approach to achieve 24/7 carbon-free computing.
\CarbonExplorer considers two important inputs: time-series data that details the power demand of large-scale datacenters and the intermittent nature of renewable energy generation at specific geographic locations (Figure~\ref{framework}-left). 
Next, \CarbonExplorer characterizes a solution space that spans the following dimensions (Figure~\ref{framework}-center):
\begin{itemize}
\item Investments in varied types of renewable energy, 
\item Investments in varied amounts of energy storage, 
\item Scheduling that shifts varied amounts of computation.
\end{itemize}
Finally, \CarbonExplorer models the datacenter design space and minimizes the carbon footprint, accounting for both operation \textbf{and} embodied carbon (Figure~\ref{framework}-right). 
%Carbon Explorer demonstrates that, depending on geographic locations, carbon-optimal strategies vary from one to another. 
%Furthermore, depending on current sustainability investments and projected renewable energy availability in the near or further future, \CarbonExplorer can guide carbon-free strategies at different time granularities to achieve carbon footprint optimality as well. 

%A holistic approach in designing a 24/7 operational carbon-free datacenter requires multiple important components: (1) data collection and evaluation at a fine, hourly granularity across power grids at different geographic locations;
%(2) fair, comparative evaluation of different alternative solutions; and (3) consideration of both operational and embodied carbon footprint in the analyses.

{\CarbonExplorer}'s inputs include two hourly time series. The first details power consumed by each of {\facebook}'s datacenters in various locations across the United States. The second details energy generation from the local grids in each datacenter location. Table~\ref{dc_renewable_investments} summarizes these locations and {\facebook}'s renewable energy investments. Section~\ref{3_dc_demand_supply} characterizes datacenter energy demand and renewable energy supply, which has implications for a production datacenter's carbon footprint. 

We evaluate three distinct solutions for 24/7 carbon-free datacenters. First, datacenters could offset their energy consumption with renewable energy generation. Operators invest in wind and solar farms on the power grids that supply their datacenters. Moreover, they implement power purchase agreements, which issue credits for renewable energy generated from those investments and offset datacenter energy consumed. This state-of-the-art solution has been central to hyperscale datacenters' pursuit of Net Zero goals. 

Second, datacenters could install energy storage and batteries to handle the intermittent availability of renewable energy. Although today's datacenters do not yet deploy batteries to manage their operational carbon footprint, they do deploy batteries to ensure system resilience and shave power peaks~\cite{Kontorinis:isca:2012, Malla:micro:2020}. As lithium-ion batteries mature, they become cost-effective for deployment at scale. On-site energy storage enables a new strategy for 24/7 carbon-free datacenters. 

Finally, datacenters could schedule computation in response to renewable energy supply. Such demand response likely requires investment in additional servers. A datacenter that defers tasks when renewable energy is scarce must compute for those tasks when renewable energy is abundant, generating demand for servers above and beyond typical loads. In effect, bursts of renewable energy generate bursts of computation and demand for servers. 

The three solutions lead to trade-offs between operational and embodied carbon footprints. Renewable energy permits carbon-free operation but is constrained by energy availability and consistency, which varies with geography. Large batteries store carbon-free energy but incur carbon overheads from manufacturing. Additional servers permit demand response and scheduling but also incur carbon overheads from manufacturing ~\cite{gupta:hpca:2021}. 

\CarbonExplorer defines a comprehensive design space for 24/7 carbon-free datacenters. Section~\ref{4_carbon_aware_dc} navigates trade-offs in the solution space with a quantitative approach. And Section~\ref{5_evaluation} illustrates the solution space for various geographic locations, highlighting the impact of site selection for future datacenters.

\section{Operational Grid Inputs: Demand and Supply Characteristics}
\label{3_dc_demand_supply}

%\begin{figure*}[t]
%\centering
%\includegraphics[width=1.4\columnwidth]{img/energy_supply_scenarios_12.pdf}
%\vspace{-0.5cm}
%\caption{Illustration of datacenter power supply scenarios: (1) Net Zero Using Offsets, (2) 24/7 Carbon Free.}
%\label{energy_supply_scenarios}
%\end{figure*}

\begin{figure*}[t]
\centering
\includegraphics[width=1.6\columnwidth]{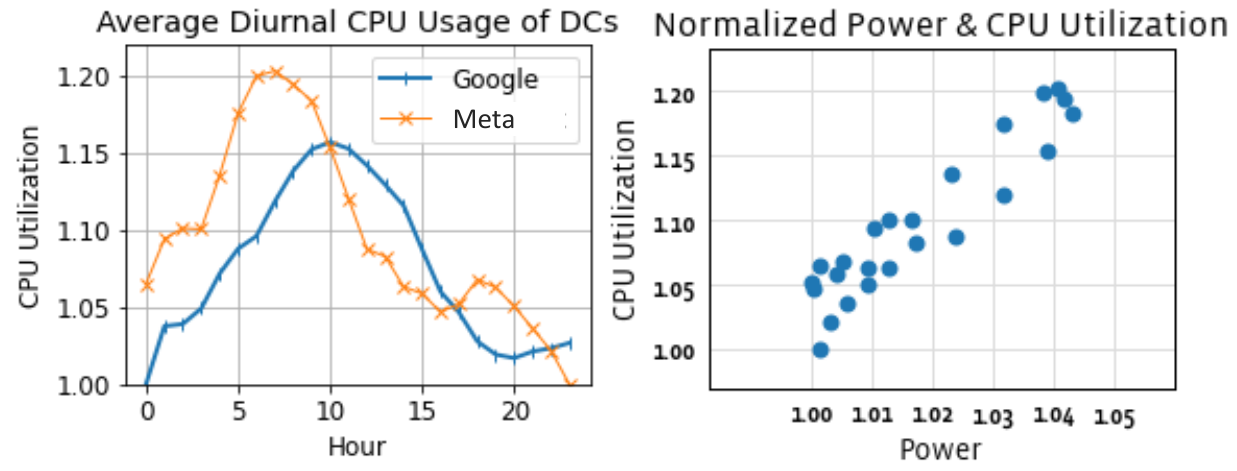}
%\vspace{-0.5cm}
\caption{[left] Hourly DC CPU fluctuations of \facebook and Google DCs.
[right] Hourly CPU Utilization and Power correlation of \facebook DCs.
}
%\vspace{-0.3cm}
\label{daily_dc_power}
\end{figure*}

Carbon-aware datacenter design requires understanding datacenter energy demand and renewable energy supply, at fine granularity and in every region. Table~\ref{dc_renewable_investments} lists the locations of {\facebook}'s datacenters and renewable energy investment, each identified by the balancing authority for the electric grid~\cite{fb_renewables}. Total investment in renewables is nearly six Gigawatts. 

\setlength{\tabcolsep}{0.2em}
\renewcommand{\arraystretch}{1.8}
\begin{table}[t]
\caption{\facebook's Datacenter Locations in the U.S. and Regional Renewable Investments~\cite{fb_renewables}}
\fontsize{7}{7}\selectfont
\label{dc_renewable_investments}
\begin{tabular}{l|c|c|c|c}
\multicolumn{1}{l|}{\multirow{2}{*}{\textbf{Location}}} & \multicolumn{1}{c|}{\multirow{2}{*}{\textbf{Balancing}}} & \multicolumn{3}{c}{\textbf{Renewable Investment [MW]}}
\\ \cline{3-5} 
\multicolumn{1}{c|}{} & \multicolumn{1}{c|}{\textbf{Authority}} & \multicolumn{1}{c|}{\textbf{Solar}} & \multicolumn{1}{c|}{\textbf{Wind}} & \multicolumn{1}{c}{\textbf{Total}}\\
\hline
1. Sarpy County, Nebraska (NE) & SWPP & 0 & 515 & 515 \\
\hline
2. Prineville, Oregon (OR)  & BPAT & 100 & 0 & 100\\
\hline
3. Eagle Mountain, Utah (UT) & PACE & 694 & 239 & 933 \\
\hline
4. Los Lunas, New Mexico (NM) & PNM & 420 & 215 & 635 \\
\hline
5. Fort Worth, Texas (TX) & ERCO & 300 & 404 & 704 \\
\hline
6. DeKalb, Illinois (IL) & PJM & & & \\
7. Henrico, Virginia (VA) & PJM & 840 & 309 & 1149 \\
8. New Albany, Ohio (OH) & PJM & & & \\
\hline
9. Forest City, North Carolina (NC) & DUK & 410 & 0 & 410  \\
\hline
10. Altoona, Iowa (IA) & MISO & 0 & 141 & 141 \\
\hline
11. Newton County, Georgia (GA) & SOCO & 425 & 0 & 425 \\
\hline
12. Gallatin, Tennessee (TN) & TVA & 742 & 0 & 742 \\
13. Huntsville, Alabama (AL) & TVA & & & \\
\hline
\multicolumn{2}{r|}{\textbf{Total}} & 1823	& 3931 &	5754 \\
\end{tabular}
\end{table}

Our energy supply analysis draws data from the U.S. Energy Information Administration (EIA) Hourly Grid Monitor, which provides operating data for power grids in the lower 48 states~\cite{eia_grid_data}. Launched in 2019, the monitor provides hourly generation statistics by collecting data from balancing authorities (BAs). Each BA operates a grid and balances electricity flows, controlling electricity generation and transmission within its own region and between neighboring authorities. 

%\CarbonExplorer projects hourly wind and solar energy supply by scaling EIA grid data in proportion to \facebook's renewable investments. It then matches hourly renewable supply against hourly datacenter demand for every region. 

\if 0 

In our power supply characteristic analysis, we draw data from the Energy Information Administration (EIA) Hourly Grid Monitor, which provides a centralized and comprehensive source for operating data for the electric power grids in the lower 48 states of the United States~\cite{eia_grid_data}. The monitor was launched in late 2019 and data is updated with new generation statistics every hour. EIA collects the data from balancing authorities (BAs) that operate the grid and are responsible for maintaining the electricity balance within its region. They achieve this balance by controlling the generation and transmission of electricity throughout its own region, and between neighboring authorities. 

Carbon Explorer projects the hourly wind and solar energy
generation from \facebook’s renewable investments by scaling EIA's grid level data in proportion of the investments of the company. Then, hourly DC power draw is mapped to the hourly renewable supply data for every region.

\fi

\subsection{Characterizing Datacenter Power Demand}

\facebook has built hyperscale datacenters across the globe with different capacities. These datacenters exhibit diurnal load patterns due to variations in user activity and exhibit peaks due to special events and holidays. Figure~\ref{daily_dc_power} shows diurnal usage for \facebook and Google datacenters and illustrates how power usage correlates with processor utilization. Google analysis is done using the open-source Borg traces~\cite{borg_traces}. {\facebook}'s CPU utilization and power is averaged over three months. For {\facebook}, CPU utilization swings by about 20\% for an average datacenter and can swing by even more for an individual datacenter. For Google, the difference between the maximum and the minimum CPU utilization is 15\%, on average~\cite{tirmazi2020borg}.

However, diurnal patterns from interactive computation do not translate directly into power patterns.
%, in part, because datacenters schedule workloads to flatten power demand. Flexible jobs are scheduled to reduce demand during peak hours and utilize servers during off-peak times. Over-provisioned servers for web-tier services can be freed during off-peak hours by up to 25\%~\cite{Tang:osdi:2020}, providing servers that opportunistically compute for delay-tolerant, batch jobs. Such schedules increase power utilization, amortizing infrastructure costs for the facility, power delivery, and servers \cite{fan2007}. Schedulers that flatten power demand has significant implications for carbon-free computing.
At datacenter scale, the difference between maximum and minimum energy demand is around 4\%, on average, which is relatively insignificant compared to the swings in renewable energy supply. Thus, in today's datacenters, power variations will arise primarily from supply but not demand. Yet shifting computation to modulate datacenter power is possible because workloads exhibit different flexibility levels and come with distinct service level objectives (SLOs). The highest priority, user-facing services require real-time response. Latency-tolerant workloads, such as batch and AI training jobs~\cite{wu2021sustainable,Tang:osdi:2020}, target specific SLO categories that include 4-, 8- and 24-hour completion times. Google has reported that flexible jobs with 24-hour completion SLOs make up about 40\% of the Borg scheduler's jobs~\cite{tirmazi2020borg}. This flexibility permits carbon-aware workload scheduling.

\subsection{Characterizing Datacenter Power Supply}

%\begin{figure}[t]
%\centering
%\includegraphics[width=0.8\columnwidth]{img/scenario1.pdf}
%\vspace{-0.5cm}
%\caption{Datacenter power demand and corresponding renewable investments to achieve Net Zero operational carbon free. }
%\vspace{-0.1cm}
%\label{energy_supply_scenarios_1}
%\end{figure}

%Figure~\ref{energy_supply_scenarios} describes
In this section, we present two scenarios that describe how datacenters could consume energy from rapidly evolving power grids.
%Renewable generation, with wind and solar, increasingly supplements or supplants traditional generation with natural gas and coal. 
In the first scenario, datacenter operators collaborate with utility providers to invest in renewable energy on the grids that power the datacenters by purchasing energy with sophisticated accounting frameworks that track renewable energy credits. This represent the state-of-the-art in reducing a datacenter's operational carbon footprint for Net Zero commitments \cite{google_sustainability, microsoft_sustainability, facebook_sustainability}.
In the second scenario, datacenter operates on renewable energy 24/7 by optimizing hourly supply and demand.

\begin{table}
\centering
\small
\renewcommand{\arraystretch}{1.5}
\caption{Carbon Efficiency of Various Energy Sources}
\fontsize{7.5}{7}\selectfont
\label{electricity_carbon_intenstiy}
\begin{tabular}{l|c | l | c}
\textbf{Type} & ~~\textbf{gCO$_2$eq/kWh~~} & \textbf{Type} & ~~\textbf{gCO$_2$eq/kWh~~} \\
\hline
Wind & 11 & Natural Gas & 490 \\
\hline
Solar & 41  & Coal & 820\\
\hline
Water & 24 & Nuclear & 12  \\
\hline
Oil & 650 & Other (Biofuels etc.)~ & 230
\end{tabular}
\vspace{-0.1cm}
\end{table}

\if 0 
\begin{figure}[ht]
\centering
\includegraphics[width=\columnwidth]{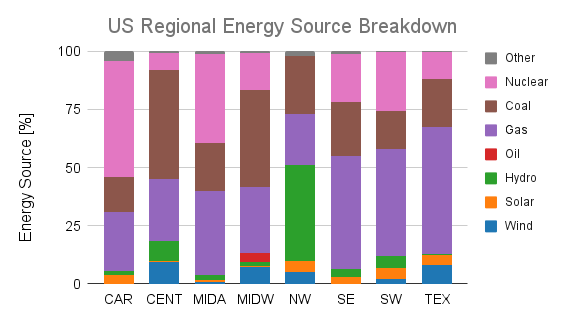}
\vspace{-0.3cm}
\caption{Energy grid mix of different regions in the US.}
\label{regional_grid_mix}
\end{figure}
\fi

\textbf{Net Zero.} Datacenter operators invest in renewable generation, such as wind and solar, and implement power purchase agreements (PPAs) to reduce datacenter exposure to the grid’s carbon intensity. PPAs link renewable energy credits (RECs) with a specific source of energy and issue, e.g., one certificate for every MWh generated~\cite{ppa_facebook, ppa_google, ppa_microsoft}. With RECs, the energy consumed is much greener than the energy offered by the grid. Table~\ref{electricity_carbon_intenstiy} details the carbon intensity of different electricity sources in the grid. The grid's energy mix is determined by the utility provider's dispatch stack and portfolio of generating assets~\cite{chase}. But the datacenter's energy mix is determined by its pre-negotiated PPAs, which deliver carbon-free energy.
%, and any additional purchases at time-of-use, which deliver carbon-intensive energy from the broader grid. Ideally, the latter is a small fraction of the total.
%Given datacenter operators' investments in renewable energy, most energy consumption may be matched and therefore is carbon-free but, during the remaining times, energy consumption is as carbon-intensive as the grid supply.

With Net Zero matching, at the end of the month (or end of the year), the total amount of energy generated and credits issued is equal or greater than the total amount of energy consumed. Although on an hourly basis, the carbon intensity of the energy used can be as much as the grid's carbon intensity during the times when there is not enough renewable supply~\cite{247_google}.

\begin{figure}[t]
\centering
\includegraphics[width=\columnwidth]{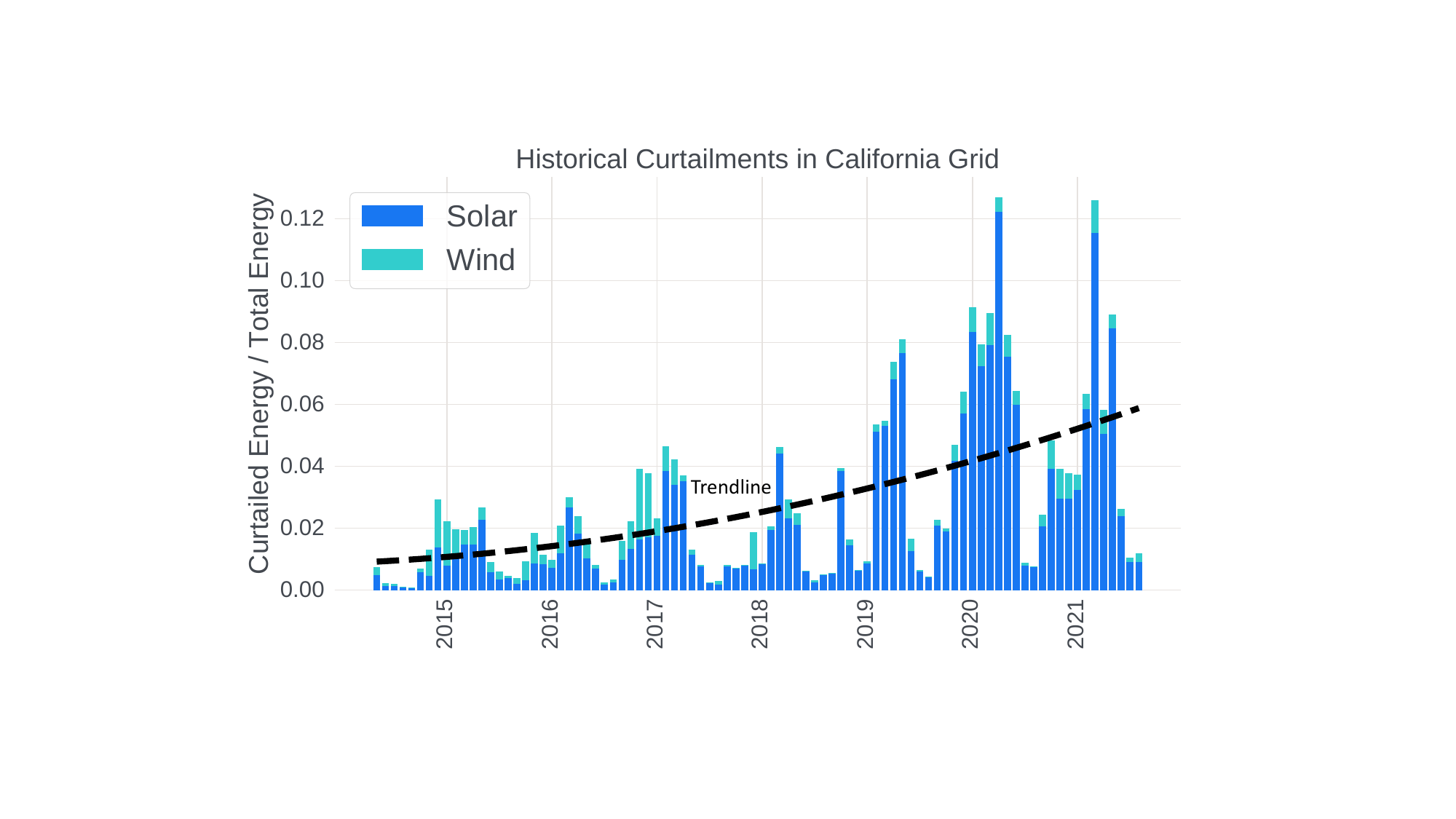}
\vspace{-0.1cm}
\caption{
%California Independent System Operator (ISO)
Wind and solar curtailments have been increasing with the renewables on the California grid~\cite{ciso_oversupply}.}
\label{curtailments}
\vspace{-0.1cm}
\end{figure}

\begin{figure*}[t]
\centering
\includegraphics[width=1.8\columnwidth]{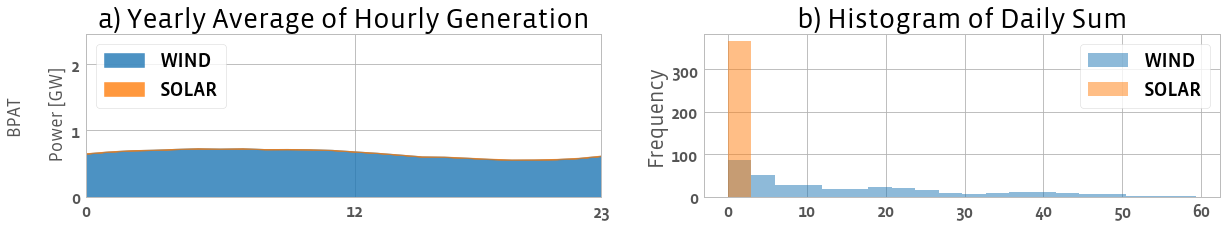}
\includegraphics[width=1.8\columnwidth]{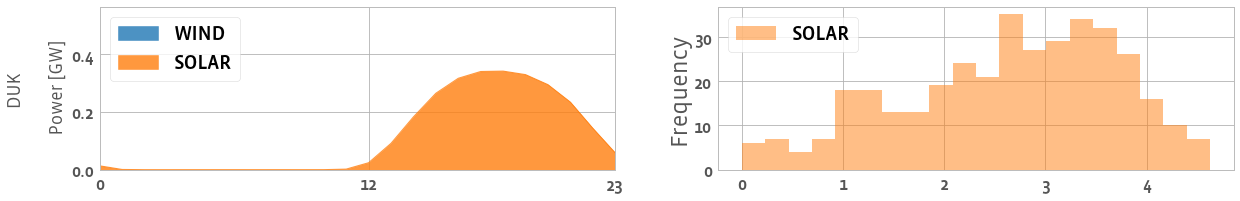}
\includegraphics[width=1.8\columnwidth]{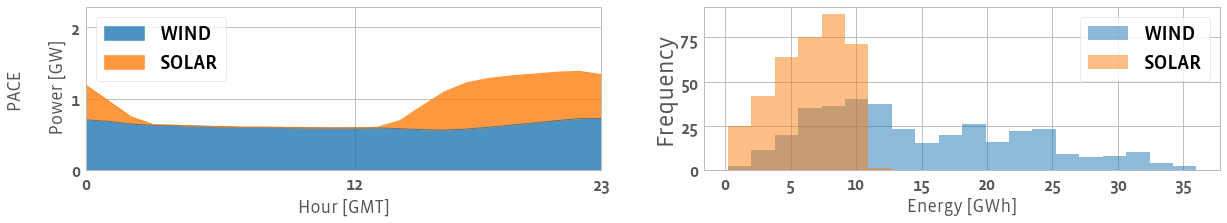}
\vspace{-0.1cm}
\caption{Figure shows hourly wind and solar generation of an average day in year 2020 (left) and histogram of daily sum to highlight day-to-day fluctuations of wind and solar generation in BPAT (in OR), DUK (in NC) and PACE (in UT) balancing authorities.
%which are composed of majorly wind, solar-only and a mix of wind and solar energy correspondingly.
The data is calculated over the entire year of 2020.}
\label{ren_seasonal_hourly}
\end{figure*}

\textbf{24/7 Carbon Free.} In addition to installing renewable energy, 24/7 carbon-free datacenters must address variable, intermittent generation. Figure~\ref{ren_seasonal_hourly} highlights variability across geography with rows corresponding to three representative regions with distinct renewable energy profiles: (a) Oregon BPAT with wind; (b) North Carolina DUK with solar; (c) Utah PACE with a mix of wind and solar. More broadly, of the ten balancing authorities in Table~\ref{dc_renewable_investments}, three offer primarily wind energy (BPAT, MISO, SWPP), three offer primarily solar energy (DUK, SOCO, TVA), and four offer a mix (ERCO, PACE, PJM, PNM).

Figure~\ref{ren_seasonal_hourly} also highlights variability across time with columns corresponding to summary statistics calculated over the year: (a) Yearly Average; (b) Histogram of Total Daily Generation~\cite{eia_grid_data}. On average, wind and solar installations provide significant supply, but averages obscure high variance across time. For BPAT, the best ten days of the year offer approximately 2.5 times more renewable energy than the average whereas the worst offer very little. Histograms quantify this variance and illustrate uncertainty in wind and solar supply.

As solar and wind farms proliferate, peaks and valleys in energy supply will become increasingly extreme. Utility providers will find it increasingly difficult to match its supply to consumer's demand. For example, California's renewable sources can generate much more electricity than needed in the middle of the day~\cite{ciso_oversupply}. And curtailments are needed to manage excess supply and reduce renewable energy generation~\cite{chien20ees, chien18aims, lin20}. Figure~\ref{curtailments} indicates that, since 2015 the curtailed gap between supply and demand has grown steadily as wind and solar capacity has increased. In 2021, curtailments reached 6\% of the total generated renewable energy in the California grid, which has deployed a significantly more renewable electricity compared to the U.S. average (33\% vs 20\% in 2020~\cite{cal_state_com, eia-renewable-supply-demand}).

It is becoming increasingly complicated to fully consume peak renewable energy generation due to the high variance. When supply exceeds demand, only generators with the lowest prices can supply energy to the grid. Prices can be zero or even negative because inputs to wind/solar farms are free and generators often receive government subsidies~\cite{zheng20, ela09}. As a result, grids may offer lower time-of-use energy prices and incentivize datacenters to defer computation to periods of abundant renewable energy.

Challenges in variable supply and curtailments require energy storage and demand response scheduling during periods of scarcity. Energy storage mitigates supply variations by providing carbon-free energy when solar and wind cannot~\cite{nrel20, doe19}. Demand response modulates datacenter demand for energy based on signals about renewable energy supply. Signals could come in the form of utility surcharges or credits when the datacenter consumes or reduces its energy demand during various times of day \cite{liu14}. Signals can also come from utility providers' generation statistics that describe the mix of green and brown energy across time and geographic locations.
%The most informative signals would communicate hourly variations in energy demand and supply.

\begin{figure}[ht]
\centering
\includegraphics[width=0.9\columnwidth]{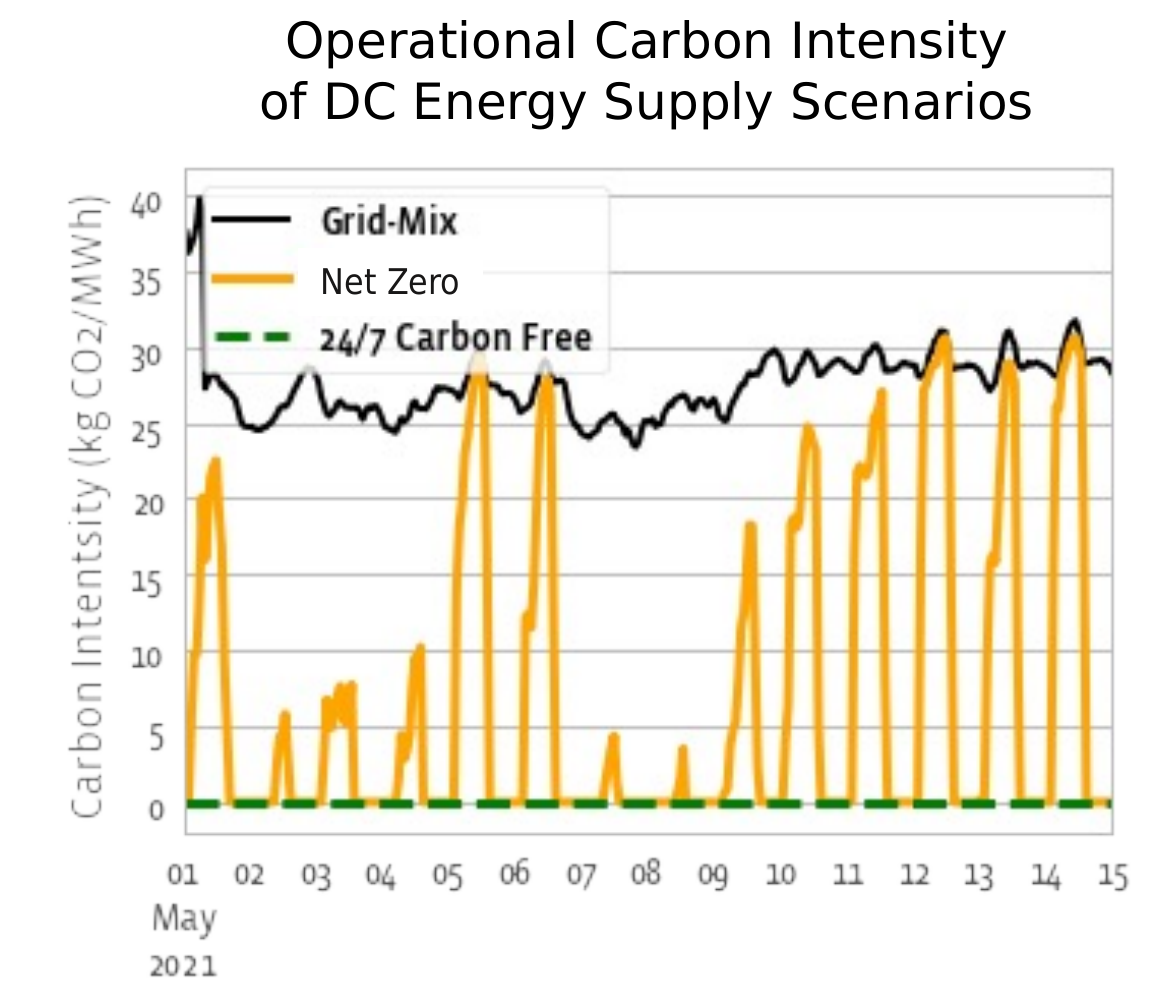}
%\vspace{-0.5cm}
\caption{Comparison of hourly operational carbon intensity of different DC energy supply scenarios.}
%\vspace{-0.1cm}
\label{energy_supply_scenarios_2}
\end{figure}

In summary, the Net Zero scenario describes how renewable energy investments significantly reduce the carbon intensity of datacenter operations. And the 24/7 scenario describes how additional investments in energy storage and demand response schedulers could further reduce carbon intensity on an hourly basis. Figure~\ref{energy_supply_scenarios_2} compares the carbon intensity of these scenarios with that of the grid's energy mix.
%Next, in Section~\ref{4_carbon_aware_dc}, we use \CarbonExplorer to show how coordinated strategies for deploying renewable energy generation, energy storage, and demand response scheduling could lead datacenters to carbon-free operations on an hourly basis.

%Next, we will show how datacenter operators that progress along these scenarios would make significant advances toward carbon-free operations on an hourly basis in Section~\ref{4_carbon_aware_dc}. Perhaps equally important, our proposed framework --- \CarbonExplorer --- enables datacenter operators to devise coordinated strategies for deploying renewable energy generation, energy storage, and demand response scheduling. 

\section{Datacenter Design: Strategies for Carbon Free Computing}
\label{4_carbon_aware_dc}
A datacenter must implement a portfolio of complementary solutions to achieve its goal of using 24/7 carbon-free energy efficiently and robustly. \CarbonExplorer considers renewable energy investments (Section~\ref{4.1_ren_investments}), energy storage installations (Section~\ref{4.2_storage}), and carbon-aware scheduling (Section~\ref{4.3_scheduling}). In this section, we model and analyze these solutions and associated trade-offs in operational and embodied carbon footprints.

\subsection{Renewable Energy}

%As we show in Section~\ref{3_dc_demand_supply}, inherent variations in solar and %wind makes powering a datacenter solely on renewable energy challenging.

\CarbonExplorer determines the solar and wind investments required for datacenters in different geographic regions to increase and achieve 100\% hourly renewable coverage. We define \textit{renewable coverage} as the percentage of hours in the year where datacenter power ($P_{DC}$) is covered by renewable power ($P_{Ren}$):
\begin{small}
\begin{alignat}{2}
\label{eq:ren_coverage}
%\text{Renewable\ Coverage}=
\left\{1 - \sum_{\text{hour}}\{P_{DC}-P_{Ren}\}/\sum_{\text{hour}}P_{DC}\right\} \times100 %\\
~~~\forall \text{hour} \in \text{DateRange} \notag
\end{alignat}
\end{small}

\begin{figure*}[t]
\centering
\includegraphics[width=1.6\columnwidth]{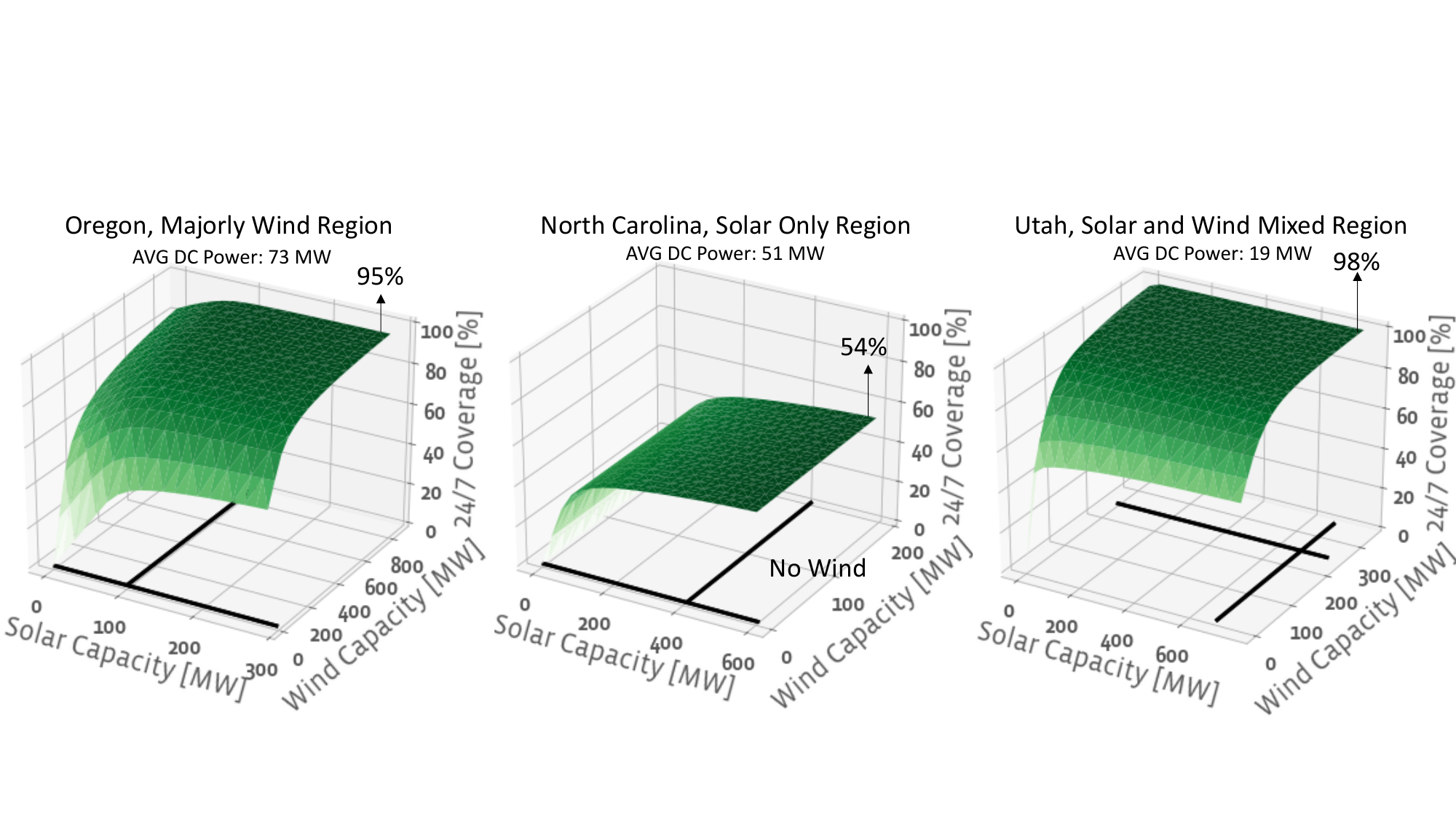}
%\vspace{-0.1cm}
\caption{24/7 coverage with varying amount of wind and solar investments. Black lines show \facebook's renewable investment
amount in the corresponding region.}
\label{24_7_coverage}
%\vspace{-0.1cm}
\end{figure*}

\begin{figure}[b]
\centering
\includegraphics[width=\columnwidth]{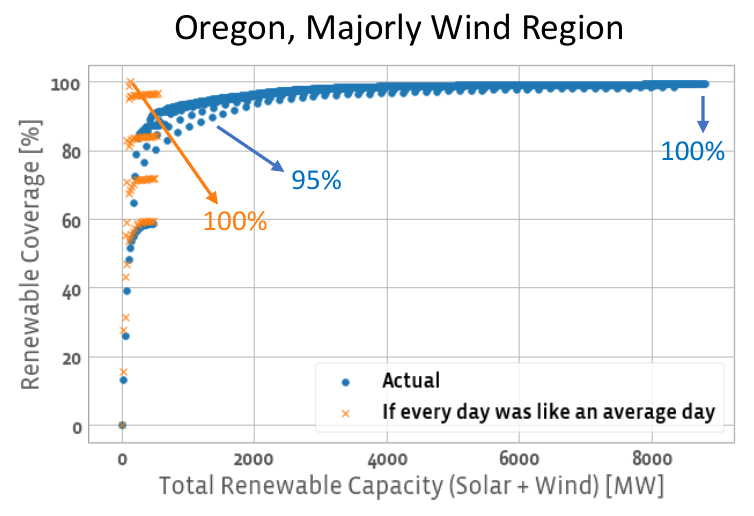}
%\vspace{-0.1cm}
\caption{24/7 coverage with different renewable investments highlighting the long tail. Each point represents a different solar + wind capacity combination.}
%\vspace{-0.1cm}
\label{24_7_coverage_bpat}
\end{figure}

\CarbonExplorer projects hourly wind and solar energy supply by scaling EIA grid data in proportion to the desired renewable investment level. It takes the maximum generated solar and wind power throughout the year as the maximum capacity of the local grid. Then, the hourly generation data is linearly scaled to the desired renewable investment capacity.
Finally, hourly renewable supply data is matched against hourly datacenter demand for every region to calculate renewable coverage. Figure~\ref{24_7_coverage} shows renewable coverage (z-axis) with different wind (x-axis) and solar (y-axis) investments from two regions served primarily by wind and solar, respectively. 

Figure~\ref{24_7_coverage} reports \facebook's existing renewable investments with black lines. While these investments help \facebook achieve Net-Zero goals on a monthly or annual basis, coverage on an hourly basis is only 46\% and 51\% in the two regions. Each region tends to favor a particular type of renewable energy generation on its local grids and \facebook's investments generally align with those profiles. One exception is Oregon, where \facebook's investments emphasize solar despite the local grid's emphasis on wind.

For regions served primarily by wind energy, like Oregon, high day-to-day fluctuations increase the investment in wind generation needed to satisfy minimum energy needs. For regions that rely entirely on solar for renewable energy, it is impossible to increase 24/7 coverage much beyond 50\% because solar energy is available only during the day. For regions that deploy a mix of solar and wind generated renewables, the tail is shorter and diminishing marginal returns in 24/7 coverage are less severe since wind and solar availability can complement each other.

\textit{There is a long tail to reach 100\% renewable coverage}. As coverage increases, curves flatten and indicate diminishing marginal returns from further investment in renewable generation. Figure~\ref{24_7_coverage_bpat} highlights the full length of the tail for Oregon's datacenter. It takes more than 5$\times$ more investments in renewable energy generation to go from 95\% to 99.9\% than to go from 0\% to 95\% coverage. Due to space limitations, we are unable to show profiles of every datacenter location. But this representative analysis shows that other solutions, such as energy storage and carbon-aware scheduling, are essential to complement renewable energy generation. 

\textit{Note that accurate hourly energy supply data is crucial when making design decisions.} Figure~\ref{24_7_coverage_bpat} shows that assuming wind and solar output to be same as average output every day leads to overly optimistic design conclusions. Under this assumption, achieving 100\% coverage would require an order of magnitude less renewable investments. Thus, \CarbonExplorer requires fine-grained time series supply and demand data when determining investments in renewable energy generation.

\label{4.1_ren_investments}

\subsection{Battery Storage}

\begin{figure}[t]
\centering
\includegraphics[width=0.9\columnwidth]{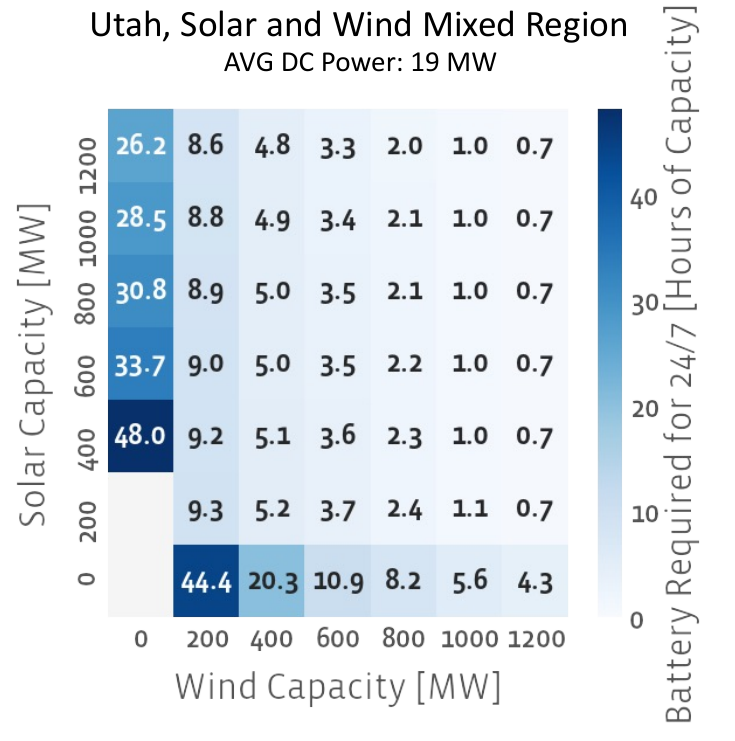}
%\vspace{-0.1cm}
\caption{How much battery needs to be deployed for 24/7 renewable energy?}
%\vspace{-0.1cm}
\label{battery_247}
\end{figure}

The Carbon Explorer framework is designed to include a modular battery model that supports different storage technologies to be added through
a simple API. In this work, we analyze lithium-ion batteries (LIB) as
improvements in energy storage over the last decade have led to LIB to
offer high capacity and energy density~\cite{diouf2015potential}. 
As the technology has matured, LIB has become a common, cost-effective storage medium for renewable energy~\cite{yang2020sustainability}. 
There is also emerging technologies such as sodium-ion (Na+) batteries, for which materials are easier to obtain and come with lower environmental impact~\cite{nayak2018lithium}.
For these reasons, batteries play an important role in 24/7 carbon-free computing. This section describes how \CarbonExplorer evaluates the impact of batteries that are charged by renewable energy and discharged by datacenter servers. 

Datacenters already deploy batteries to prevent the interruption of services during maintenance or power failures. Batteries distributed throughout racks and clusters permit continuous operation when utility power fails and the datacenter must switch to diesel generator backups~\cite{malla2020coordinated}. 

We envision batteries deployed on-site with the datacenter to reduce its carbon footprint. Batteries will be charged when there is excess renewable supply (i.e. when the amount of energy produced by the renewable deployment is larger than datacenter's demand). Batteries will be discharged to power the datacenter when there is a lack of renewable supply (i.e. when the amount of energy produced by the renewable deployment is smaller than datacenter's demand). 

The battery model used in \CarbonExplorer is the C/L/C model~\cite{kazhamiaka2019tractable}; it explicitly models several characteristics of lithium-ion batteries, including energy content limits, efficiency loss, limits on the applied power with respect to the energy content, and
linear charging/discharging rates with respect to battery capacity on minute basis.
The model also includes a parameter for controlling the depth of discharge (DoD).
Model parameters are tuned to represent a battery composed of Lithium Iron Phosphate cells (LFP)~\cite{lifecell} -- a cell type found often in large stationary storage applications.
 
Figure~\ref{battery_247} shows the amount of energy storage capacity required to reach 24/7 renewable energy coverage at different solar and wind capacities for Utah datacenter. Capacity is reported in terms of computational hours (e.g., 2 hours for a 20MW datacenter corresponds to 40MWh of battery capacity).
Regions with mixed solar and wind generation exhibit less variable, day-to-day fluctuations and can achieve 24/7 carbon-free compute with less battery capacity. By adding around five hours of battery capacity to its existing renewable investments in the region, \facebook's Utah datacenter can reach 24/7 carbon-free operational energy.
A battery of this size would be comparable to a utility-scale battery; the largest utility-scale energy storage project so far can offer 300 MW of power and 1,200-MWh of capacity~\cite{pge_battery}.

In contrast, battery capacity requirements for 24/7 are greatest for regions that rely majorly on wind. Oregon suffers from extremely high day-to-day fluctuations and there are days with almost no wind power. Requirements are also high for regions that rely entirely on solar. For example, North Carolina datacenter requires 14 hours of battery-based compute. 

\label{4.2_storage}

\subsection{Carbon Aware Scheduling}
%\begin{figure}[ht]
%\centering
%\includegraphics[width=0.8\columnwidth]{img/s2_247_coverage.png}
%\vspace{-0.3cm}
%\caption{Scenario 2: 24/7 coverage.}
%\label{s1_cas}
%\end{figure}

Carbon-aware scheduling (CAS) exploits delay tolerant workloads to achieve 24/7 carbon-free computing, shifting workloads from times when the carbon intensity of electricity sources is high to times when it is low. Hyperscale datacenter workloads are commonly organized into tiers based on their Service Level Agreements (SLAs). Higher tier jobs are latency sensitive and require high availability. 

On the other hand, lower tiers can tolerate delays. Examples temporally flexible workloads include AI model training, data processing pipelines, and offline video processing. Google traces indicate a significant fraction of jobs submitted to the Borg scheduler are in the free and best-effort-batch tiers with weak SLAs~\cite{tirmazi2020borg}. Jobs at \facebook exhibit similar characteristics.
For example, the offline data processing workloads constitute about 7.5\% of all the workloads in the fleet. Of these, about 87.4\% of the workloads have SLOs that are greater than 4-hours with a majority having 24-hour SLOs. This provides great flexibility in workload time shifting to optimize carbon.

\begin{figure}[h]
    \centering
    \includegraphics[width=1\columnwidth]{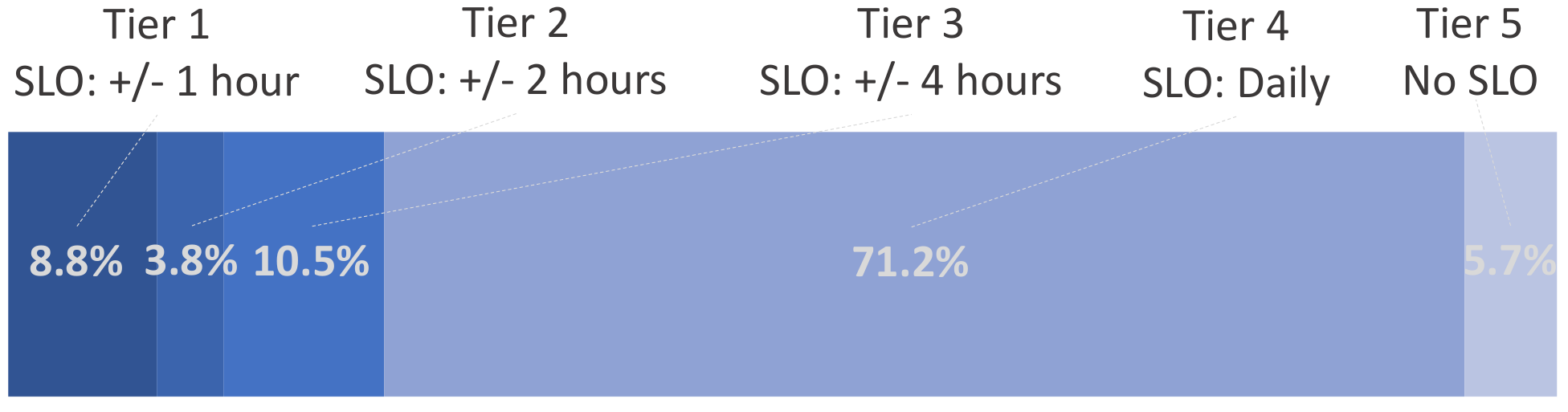}
    \vspace{-0.5cm}
    \caption{Breakdown of data processing workloads by completion time SLO at} \facebook.
    \label{fig:dataswarm_tiers}
\end{figure}

\begin{figure}[t]
\centering
\includegraphics[width=0.9\columnwidth]{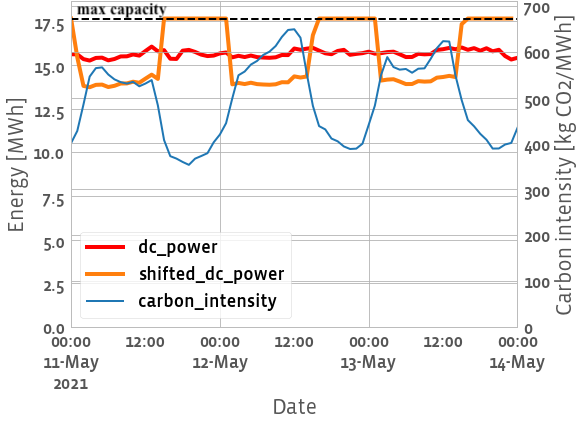}
\vspace{-0.1cm}
\caption{Carbon aware scheduling illustration for the Utah DC.}
%\vspace{-0.1cm}
\label{s1_cas}
\end{figure}

\CarbonExplorer estimates the potential benefits of carbon aware workload scheduling using a greedy algorithm. The algorithm takes two customizable input constraints: datacenter capacity and flexible workload ratio for each hour of the day. Given these two constraints, flexible workloads are moved from times of highest carbon intensity to times of lowest intensity until all flexible workloads have been moved or all datacenter servers have been used for the given hour. 

\begin{algorithm}[ht]
\small
\textbf{Input Constraint 1:} $P_{DC_{MAX}} = Maximum \ DC\ Capacity$\\
\textbf{Input Constraint 2:} $FWR = Flexible\ Workload\ Ratio\ (\%)$ \\
\textbf{Goal:} For each day, minimize:
$$\sum_{h \in hour} \{P_{DC}(h)-P_{Ren}(h)\} $$ \\
where \\
\Indp
$P_{DC}(h) < P_{DC_{MAX}}$ and \\
$P_{DC}(h) \times FWR$ is allowed to shift\\
\Indm
%\textbf{Algorithm:}\\
%\For{day in Date\Range}
%{
%\While{ $P_{DC} < P_{DC_{MAX}}$}\\
%{
% $min(P_{DC}) += P_{DC}(h) \times FWR$ \\
% $max(P_{DC}) += P_{DC}(h) \times FWR$ \\
%}
%and \\
%}
\end{algorithm}

\begin{figure}[t]
\centering
\includegraphics[width=0.9\columnwidth]{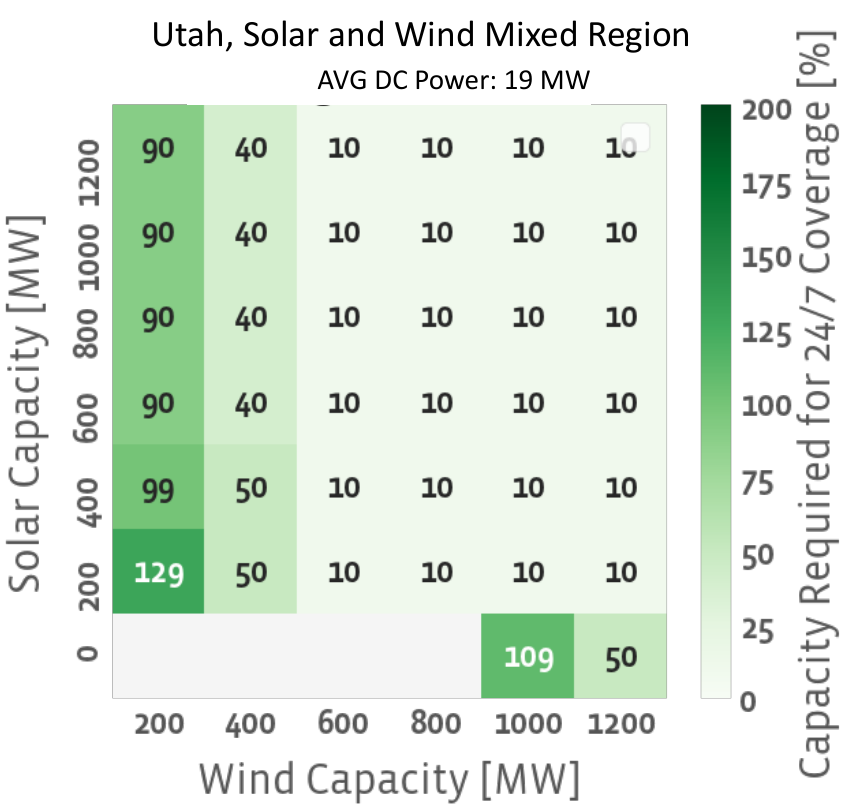}
%\vspace{-0.1cm}
\caption{Scenario 3: Carbon aware scheduling for 24/7.}
%\vspace{-0.1cm}
\label{s3_cas}
\end{figure}

%FB data warehouse tier system
%The quotas are:
%Up to 5% of the overall workload may be in tier 1
%Up to 20% of the namespace workload may be in tier 1 or tier 2
%Up to 40% of the scope or namespace workload may be in tier 1, 2, or 3
%In other words, for a namespace, usage in tiers 1 and 2 should be below 20% of the namespace’s %workload. For a scope, combined usage in tiers 1-3 should be below 40% of scope’s workload.

%\begin{figure}[ht]
%\centering
%\includegraphics[width=0.7\columnwidth]{img/cas_scenario_1/pace.png}
%%\vspace{-0.1cm}
%\caption{Carbon aware scheduling for the Utah DC with grid mix.}
%\label{s1_cas_results}
%%\vspace{-0.1cm}
%\end{figure}

Figure~\ref{s1_cas} illustrates an example of a carbon aware scheduling over three days. The blue line shows how the grid's carbon intensity varies depending on the hour of the day. The red and orange lines shows datacenter power draw when carbon aware scheduling is and is not applied. In this example, the maximum allowed power capacity of the DC is assumed to be 17.6 MW and 10\% of the workloads running every hour are flexible to finish within a day.

%Carbon aware scheduling can be used to help achieve 24/7 renewable by shifting workloads from times when there is lack of renewable energy to times with excess supply.

\textbf{Additional Capacity.} Shifting computation across time may require additional server
capacity for sustained increases in computation when carbon-free/low-carbon energy is abundant.
The need for surplus capacity reveals an interesting trade-off between operational and embodied carbon, which has not been considered by any existing work in the literature~\cite{bashir2021enabling,radovanovic2021carbon, wierman14}.

From an operations perspective, increasing the number of provisioned servers mitigates the data center’s carbon footprint by permitting demand response and reducing the carbon intensity of its energy. However, from the embodied perspective, over-provisioned servers increase embodied carbon emissions from hardware manufacturing~\cite{gupta:hpca:2021}. Therefore, there is a fine balance between operational and capital expenditures.

%required to ensure modulating data center utilization modulates demand for power. Both of these requirements are satisfied by hyperscale data centers. 

Energy-proportional design is essential when over-provisioning datacenter servers \cite{barroso2007case}. Idle servers should draw little power, especially since hourly scheduling decisions provide ample time for servers to switch between power models. Indeed, server power can be accurately modeled as a linear function of utilization with the y-intercept denoting a server’s idle power. Figure~\ref{daily_dc_power} illustrates energy-proportionality and correlation between {\facebook}'s datacenter power and its CPU utilization. 

Figure~\ref{s3_cas} shows how much server capacity is required to achieve 24/7 carbon-free computation. Additional capacity is measured as a percentage of the datacenter's existing capacity. In this example, all workloads are assumed to be flexible to shift. Analysis shows that the additional capacity required to reach to 24/7 varies between 19\% to over 100\% (i.e. doubling the number of servers). Note that, as an alternative to deploying more servers, datacenters might Turbo Boost their current servers to increase compute throughput without increasing capital costs and embodied carbon.

\if 0 

In other words, construction's carbon footprint is only 16\% of hardware's. 

\fi

%in terms of percentage of the existing server capacity, is required to achieve 24/7 carbon free computing.We define capacity in terms of datacenter power (\textit{e.g.}, 1MW of spare capacity in a 20MW datacenter). Any extra capacity needed beyond the maximum power is counted as additional capacity required.

\label{4.3_scheduling}

\section{Carbon Minimization: Holistic Design Exploration}
\label{5_evaluation}

\begin{figure}[t]
\centering
\includegraphics[width=0.9\columnwidth]{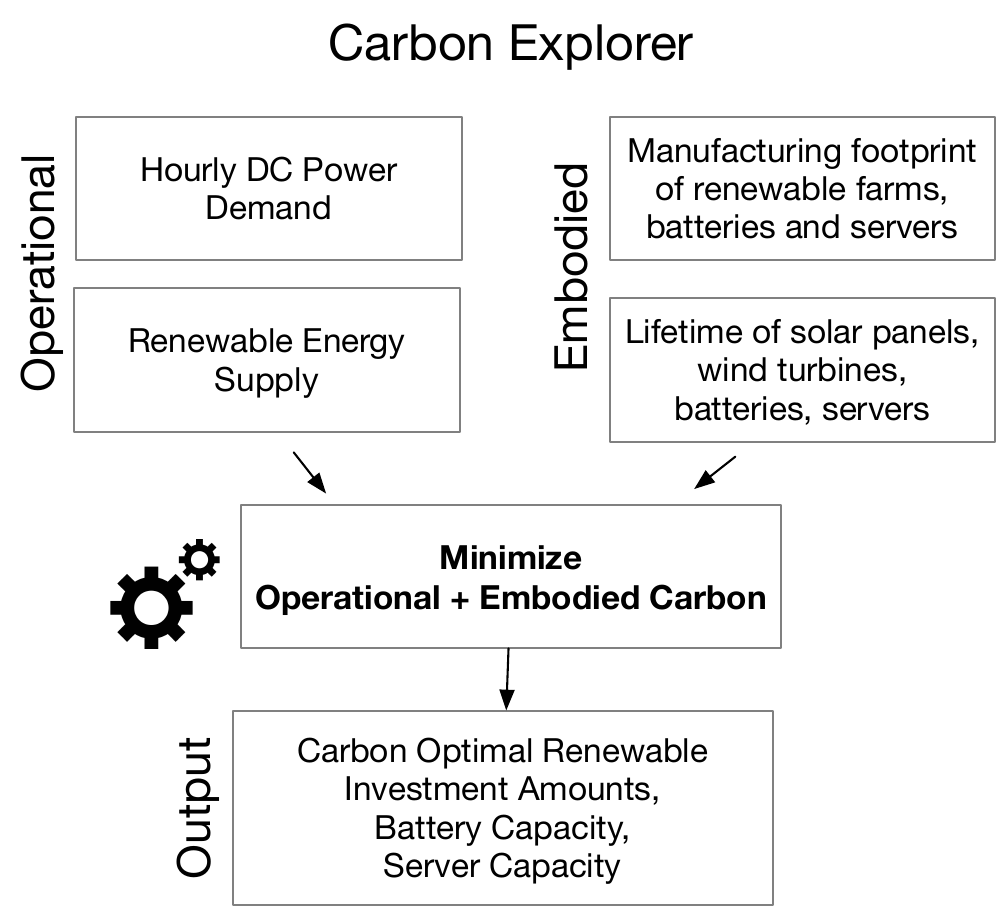}
\caption{Carbon Explorer}
\label{carbon_explorer_process}
\end{figure}

%\begin{figure*}[t]
%\centering
%\includegraphics[width=1.8\columnwidth]{img/combined/combined.pdf}
%\caption{Operational and embodied footprint of the three solutions. Pareto frontier shows how the long tail to reach 100\% renewable coverage can be shortened with complementary solutions.}
%\label{combined}
%\end{figure*}

\begin{figure}[!htb]
\centering
\includegraphics[width=1\columnwidth]{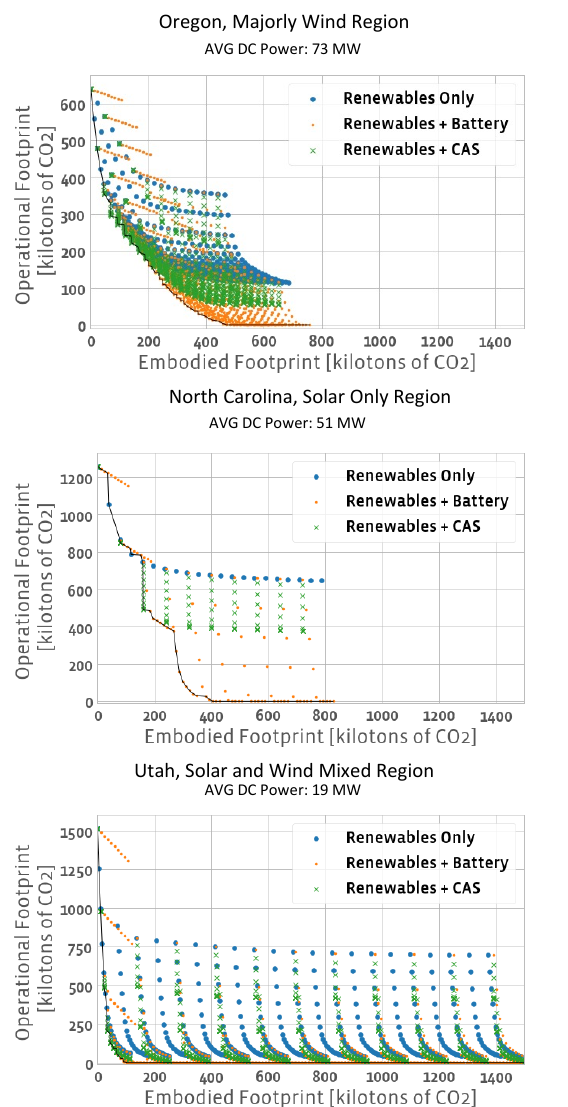}
\caption{Operational and embodied footprint of the three solutions. Pareto frontier shows how the long tail to reach 100\% renewable coverage can be shortened with complementary solutions.}
\label{combined}
\end{figure}

% In this section we show how \textit{Carbon Explorer}'s operational and embodied carbon footprint analysis helps evaluate solutions in the design space.

Reaching 24/7 carbon-free computing comes with non-negligible embodied carbon costs. Thus, \CarbonExplorer must consider \textit{both} operational and embodied carbon when minimizing the overall carbon footprint. Figure~\ref{carbon_explorer_process} presents the process of identifying an optimal datacenter design point from the carbon footprint's perspective.

First, \CarbonExplorer requires inputs for its models of operational and embodied carbon. Operational inputs include hourly datacenter power demand and renewable power supply. Embodied inputs account for the carbon emissions from manufacturing and the expected lifetimes of solar and wind farms, lithium-ion batteries, and datacenter servers. 

\CarbonExplorer exhaustively searches the design space to minimize the sum of operational and embodied carbon. The design space includes the three solutions --- renewable, battery, server investments --- described in detailed in Section~\ref{4_carbon_aware_dc}. Datacenter operators specify the bounds of the design space. Finally, \CarbonExplorer outputs the carbon-optimal investments in renewable energy generation, battery capacity, and server capacity.

\subsection{Embodied Carbon}

\textbf{Renewables.} The manufacturing (or embodied) carbon footprint for wind turbines ranges from 10-15 grams of \carbon per kWh whereas the footprint for solar farms ranges from 40-70 grams of \carbon per kWh~\cite{ren_life_cycle}. These numbers are derived from a life cycle analysis and accounts for manufacturing costs and the expected amount of energy generated over the asset's lifetime. The average lifetime for solar panels is 25-30 years and that for wind turbines is 20 years.

\textbf{Batteries.} The manufacturing footprint of lithium-ion batteries ranges from
74 to 134 kilograms of \carbon per KWh of battery capacity~\cite{emilsson2019lithium, battery_footprint} and comprised of three major steps.
(1) The majority of this footprint comes from production of the upstream battery materials,
which is 59 kg per KWh and 44-80\% of the total footprint.
(2) Cell production and assembly is the part which uses the most electricity and hence the
footprint ranges from 0-60 kg per KWh and is 0-44\% of the total footprint depending on renewable energy usage during production.
(3) End-of-life processing/recycling, which is a necessary and challenging task for batteries, generates 15 kg per KWh and is 11-20\% of the total footprint~\cite{yang2020sustainability}.
The lifetime of the battery is calculated in terms of the number of discharge cycles. Utility-scale batteries, such as Tesla's Powerpack, last 3000-4000 cycles~\cite{batter_cycles}.

Lithium Iron Phosphate technology we model in this paper allows a large number of 
charge/discharge cycles. Charging rate (C-Rate) and Depth of Discharge are two of the 
important factors effecting the lifetime of the battery. In this work, we assume C-Rate of 1C meaning full discharge or charge in one hour as our renewable energy data is on hourly basis.
DoD provides a maximum usage limit over batter's total energy capacity, to extend it's lifespan.
In standard environment with proper temperature, humidity, and for 1C-rate cycles, life cycle estimation for LFP batteries are 3000 cycles at 100\% DoD, and 4500 cycles at 80\% DoD~\cite{lfp_lifetime}. We studied these two scenarios.

\textbf{Servers.} The manufacturing footprint of servers is estimated to be 744.5 kg eq \carbon~\cite{hpe_footprint}, using an HPE ProLiant DL360 Gen10 server as a proxy. This server includes a single-socket CPU with 48 GB DRAM and has Thermal Design Power (TDP) of 85 Watts. Carbon measurements include its mainboard, SSD, daughterboard, enclosure, fans, transport and assembly. We estimate server lifetime of five years. 

When demand response requires additional servers,
to capture additional cost for floor-space and other infrastructure required,
we added a parameterized surcharge cost.
Even if new facility construction were required, initial estimates suggest carbon from construction is small compared to carbon from servers. 
For example, of Meta's 2019 Scope 3 carbon emissions, 42\% and 7\% is attributed to hardware infrastructure and construction, respectively~\cite{gupta:hpca:2021}. In other words, construction's carbon footprint is only 16\% of hardware's. This large difference might be explained by the relatively longevity of existing datacenters and rarity of new construction. A hyperscale datacenter's lifetime is 15 to 20 years whereas server hardware is typically three to five years. Therefore, Carbon Explorer models total embodied carbon by applying the multiplier to be 1.16x to server's embodied carbon.

However, sustainability reports carry several caveats. Hardware's carbon accounts for both refreshing existing datacenters and provisioning new ones. Construction's carbon is reported in the year incurred rather than amortized. Reports do not breakdown carbon incurred by a datacenter and its hardware. Further analysis will require better public data and carbon models for construction.

%\subsection{Carbon-Efficient 24/7 Carbon Free Datacenters: A Holistic Analysis}
%\subsection{24/7 Carbon Free Datacenters: \\ A Holistic Analysis}
\subsection{A Holistic Analysis}

\begin{figure*}[t]
\centering
\includegraphics[width=2\columnwidth]{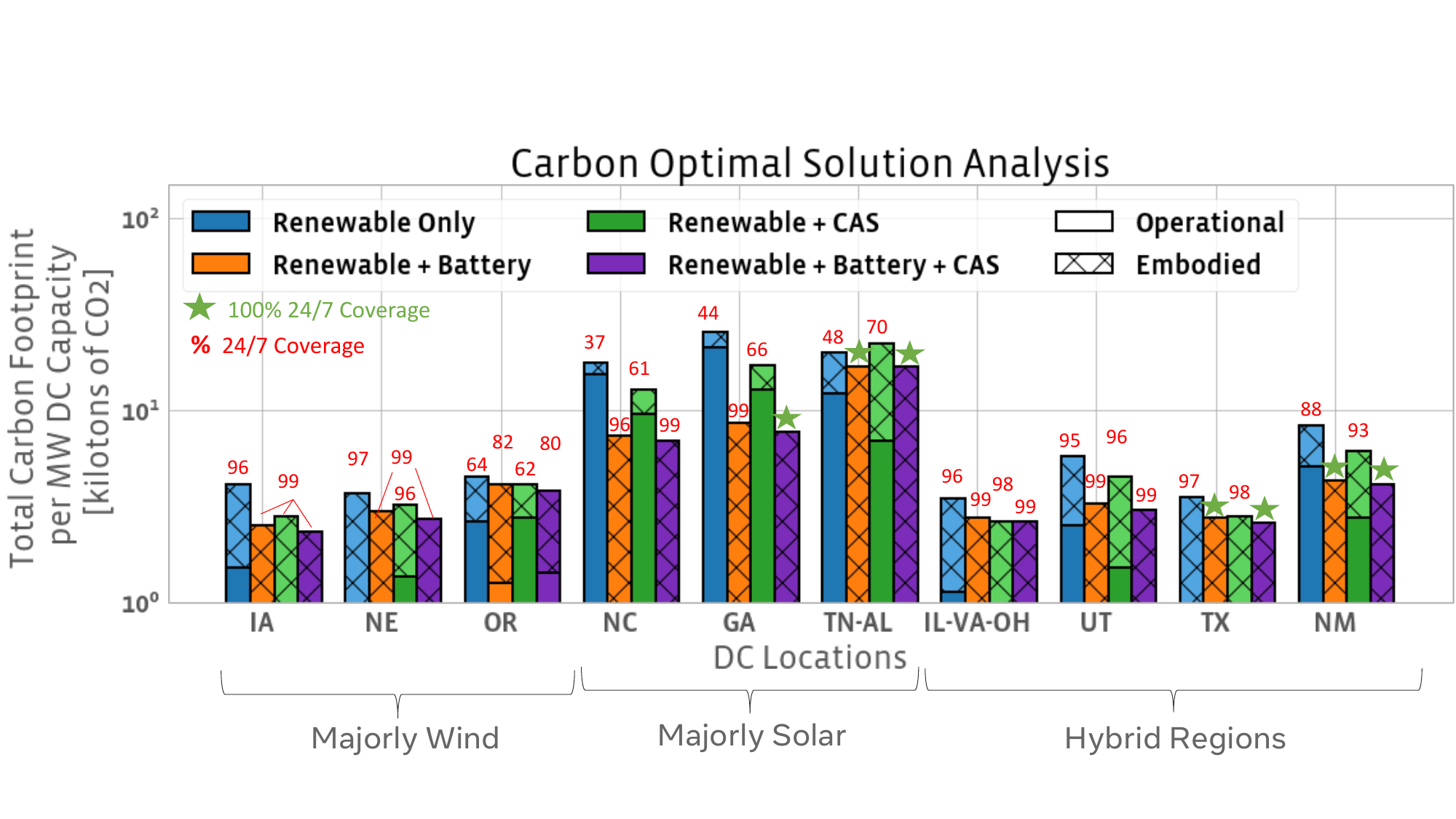}
\vspace{-0.4cm}
\caption{The total footprint of the carbon-optimal setting of each solution per MW DC capacity.}
\label{total_all}
\end{figure*}

Figure~\ref{combined} illustrates unique trade-offs between decreases in operational carbon (y-axis) and increases in embodied carbon (x-axis). In this evaluation, we assume 40\% of datacenter workloads are delay-tolerant, a realistic flexible workload ratio \cite{tirmazi2020borg}, and can be deferred for carbon-aware scheduling. We examine four strategies: renewable energy generation alone, renewables with batteries, renewables with carbon-aware scheduling, and renewables with batteries and carbon-aware scheduling (Section~\ref{4_carbon_aware_dc}). 

The space includes solutions that can significantly reduce a datacenter's overall carbon footprint. Reductions in operational carbon incur increases in embodied carbon.
Datacenter operators must be careful in its pursuit of 24/7 coverage because some solutions incur much higher embodied carbon costs than others. Renewable generation alone is insufficient and solutions that combine renewable energy generation with batteries and scheduling are needed. 

The Pareto frontier indicates that any solution for 24/7 carbon-free operations (i.e., zero operational carbon) must include renewable energy and batteries. Moreover, as 24/7 coverage increases, solutions that include batteries will incur smaller embodied carbon costs than solutions that rely solely on renewable energy and/or deploying additional servers to support carbon-aware scheduling. 

Unfortunately, the Pareto frontier exhibits a long tail, which indicates increasingly expensive solutions required to reach full 24/7 coverage. For example, in Oregon, investments in renewable energy and batteries can quickly reduce operational \carbon from ~600 to ~400 kilotons with minimal embodied cost. But eliminating the remaining 400 kilotons of operational carbon will require significantly larger embodied carbon cost.

Figure~\ref{total_all} details the most effective strategy for 24/7 carbon-free datacenter operation by geographic location and availability of renewable energy. We show the total carbon footprint, breaking down operational (solid) and embodied (cross-pattern) components. The carbon footprint is normalized relative to datacenter sizes, measured in MW of power capacity. We annotate each bar, identifying solutions that achieve full 100\% 24/7 coverage (green stars) and those that make partial progress (red percentages). 

\textbf{Renewables Only.} Relying solely on renewable energy generation incurs the highest embodied carbon costs in every geographic region. The 24/7 renewable coverage with renewables only ranges from 37\% to 97\% depending on geographical regions. Because renewable energy generation is intermittent, datacenter operators would need a large number of solar or wind farms to ensure sufficient supply during supply valleys. Even with significant investment, however, renewable energy supply fundamentally depends on weather and time of day, which leads to incomplete 24/7 coverage and higher operational carbon footprints that dominate the datacenter's total carbon footprint. 

Figure~\ref{total_all} indicates the most effective renewable solutions include wind farms. A combination of wind and solar farms provides complementary generating assets and mitigates supply variance. Hybrid geographic regions, which use both wind and solar, achieve higher 24/7 coverage that ranges from 88\% to 97\%. 

Regions that rely primarily on wind can also achieve high 24/7 coverage with careful datacenter site selection. Out of the thirteen locations, datacenters in Nebraska (a majorly-wind region), Utah and Texas (wind and solar hybrid regions) stand out as the best locations to minimize total carbon footprint and achieve highest coverage. Valleys in energy supply are shallower in these windy regions compared to those in others. In contrast, regions that rely primarily on solar (i.e., NC, GA, TN, AL) struggle to achieve full 24/7 coverage and incur the highest carbon footprints since solar energy is only available during parts of the day.  

%Out of the thirteen locations, Nebraska (a majorly-wind region), Utah and Texas (wind and solar hybrid regions) datacenters stand out as the best locations to minimize total carbon footprint and achieve highest coverage. The lowest energy generation days throughout the year in these regions has higher energy generated compared to other regions.

\textbf{Renewables + Battery.} The addition of batteries reduces the total carbon footprint by an order of magnitude in all geographic regions. The reduction is most pronounced in regions that rely only on solar energy. For four of the thirteen datacenter regions (TN, AL, TX, NM), the most carbon-efficient solution deploys enough battery capacity to achieve 100\% 24/7 coverage and completely eliminate the datacenter's operational carbon footprint. For the other regions optimal renewable energy coverage ranges from 82\% to 99\%.
The battery capacity amount to achieve the optimal footprint ranges from 200MWh - 1800MWh for different datacenters and it would represent a utility scale battery capacity.
Given hyperscale datacenters cost billions of dollars~\cite{google_cost}, this battery investment represents a small fraction of a data center's overall cost at current battery prices of \$350/kWh~\cite{cole2021cost}. 

\begin{figure}
\centering
\includegraphics[width=1\columnwidth]{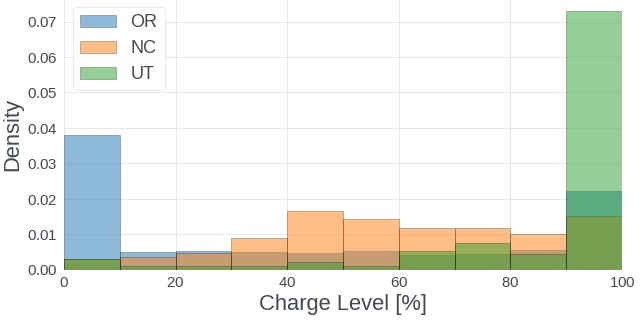}
\vspace{-0.5cm}
\caption{Battery charge level distributions}
\label{bat_charge_dist}
\end{figure}

Suppose we manage batteries such that the depth of discharge (DoD) is 100\%. Under the carbon-optimal battery configuration, Figure~\ref{bat_charge_dist} indicates batteries are often fully charged or fully discharged. Although this outcome arises naturally from our scheduling algorithm, which maximizes the battery usage to avoid carbon-intensive energy, high DoD can harm battery lifespan. Conversely, lower DoD may benefit lifespan but reduce the batteries' usable capacities, requiring larger batteries or using carbon-intensive energy. Thus, we discover a trade-off between embodied and operational carbon. 

For example, compare DoD of 100\% and 80\%. The lower DoD of 80\% increases battery lifespan and the number of (dis)charge cycles by 50\%. But because shallower discharges reduce the effective capacity of the battery, larger batteries are required and the associated embodied carbon increases by 43\% in the carbon-optimal configuration. Evaluating the net impact, Carbon Explorer finds that 80\% DoD lowers total carbon by 5\% on average, across geographic regions.  

Lowering DoD further increases battery lifetime but requires even larger batteries~\cite{lfp_lifetime}. At some point, shallower DoD becomes counterproductive. A 60\% DoD implies 10,000 daily (dis)charge cycles and 27-year battery lifespan. Other degradation factors would come in to play before reaching the 27-year lifespan. Overall, tuning DoD can lower total carbon by 3-9\% across the DC regions.

\if 0 
\hl{In the optimal battery capacity levels with 100\% depth of discharge, the charge level distribution of the batteries exhibit peaks at the lowest and highest charge levels for majority of the regions. This is expected given our algorithm aims to maximize the battery utility however this may create a concern in terms of battery lifespan as high DoD levels can compromise the lifespan. Lower DoD levels increase the lifespan however they reduce the usable capacity of the battery and this may create a trade-off between operational and embodied footprint. To evaluate this trade-off, we compared the 100\% DoD with 80\% DoD. 80\% DoD increases the lifetime by 50\% cycles however since it reduces the effective capacity of the battery, found that it requires 43\% higher embodied footprint in the carbon optimal scenario. Overall in terms of the total footprint, 80\% DoD results in 5\% lower carbon footprint on average of different regions. Further lowering DoD can increase the lifetime more according to the models that use multiple cycles per day}
\fi

\if 0 
\hl{When used for renewable energy storage purposes,
60\% DoD would imply 27 years of battery lifetime at 10000 cycles with 1 cycle per day.
In this setting, other degradation factors would come in to play before reaching the 27 years of lifespan.
Overall, tuning DoD can reduce the battery in the range of 3-9\% depending on the region and
5\% on average.}
\fi

\textbf{Renewables + CAS.} Carbon-aware scheduling provides an alternative to batteries, increasing 24/7 coverage by 1\% to 21\% across geographic regions. \CarbonExplorer finds that deploying 6\% to 76\% additional server capacity allows the scheduler to move computation from periods when renewable energy is scarce to periods when it is abundant, thereby reducing the datacenter's overall carbon footprint. However, carbon-aware scheduling is constrained by the degree of workload flexibility and the number of provisioned servers available to process deferred jobs. Due to these constraints, scheduling alone is insufficient for full 24/7 coverage in regions characterized by many days with near zero renewable energy (e.g., wind in Oregon) or regions that rely exclusively on solar energy. 

%On the other hand, since \textit{Renewables + CAS} is limited by the degree of workload scheduling flexibility and relies on additional, available server capacities, for regions like OR where there are significant number of days with near zero wind generation (see Figure~\ref{ren_seasonal_hourly}) and for solar only regions, \textit{Renewables + CAS} alone is not sufficient to reach 24/7.

\textbf{Renewables + Battery + CAS.} A combination of battery deployments and carbon-aware scheduling offers additional improvements. We use a heuristic based solution where the priority is given to the workloads to minimize the runtime delays. Whenever there is lack of renewable supply, the energy stored in the battery is used first and workload shifting happens only if the energy stored in the batteries are not sufficient (at maximum DoD level). Whenever there is extra renewable supply, all available workloads are executed to use the available power first and batteries are charged with the remaining supply.
This reduces the additional capacity required to reach 24/7 coverage compared with a battery only solution or a CAS-only solution alone and makes 100\% coverage the carbon-optimal solution in five of the regions. For rest of the regions except OR, carbon-optimal solution is at above 99\%.

In summary, 24/7 coverage depends on renewable energy generation characteristics of the datacenter region, delay tolerance characteristics of the workloads and the possibility of using batteries. In several cases, achieving complete 24/7 coverage is neither feasible nor the most carbon-efficient solution. Solutions that complement investments in renewable energy are necessary. The addition of batteries and carbon-aware scheduling can reduce a datacenter's total carbon footprint by 15-65\% depending on the region. However, there are other side-effects of using batteries and additional servers such as environmental factors and construction limitations. When designing carbon-aware solutions one must take the other factors into account as well.

Looking forward, to achieve a sustainable future for datacenters, understanding the design dimensions of carbon-free datacenter design is crucial. We expect the delay tolerance nature of computing to increase. For example, recent research on carbon-aware datacenter software for sustainability has started to emerge~\cite{anderson2022treehouse}. Delay-tolerant computing, such as AI training, is becoming more prevalent as well~\cite{wu2022sustainable}. At the same time, the strong demand for a clean energy future has propelled significant efficiency improvement for both renewable infrastructures and energy storage technologies~\cite{firstsolar, canadiansolar}. Thus, in addition to computing's delay tolerance, a carbon-optimal datacenter design will also hinge on the trade-off between renewable energy utilization (as a result of operational efficiencies of energy generation and storage) and the associated embodied carbon cost. This is exactly the design space and optimization that Carbon Explorer enables.

\section{Discussion and Related Work}
\label{related_work}

\textbf{Renewable Energy.} Prior academic research emphasizes renewable energy on-site at the datacenter \cite{li2012isca}. Computation uses local, solar energy and minimizes energy consumed from the grid \cite{goiri13, goiri2014}. The datacenter's power infrastructure is enhanced to switch between multiple types of local generators and microgrids \cite{liu16ics, li2013hpca, li2012isca}. And strategies are developed to deploy scale out servers and renewable generators in a modular fashion \cite{li13micro, li16}. These strategies seem sensible for edge and fog servers \cite{li2015isca}. However, the hyperscale datacenters we study avoid many of these challenges. They do not need to manage local power generation because they have invested in renewable generation on the grid at scales that are unlikely on-site. Yet they improve sustainability through power purchase agreements. We study renewable energy across geographic locations and coordinate their installation with battery and server provisioning at scale. 

% HEB: Deploying and managing hybrid energy buffers for improving datacenter efficiency and economy. Liu et al. ISCA-2015.
% Exploring highly dependable and efficient datacenter power system using hybrid and hierarchical energy buffers Transactions on Sustainable Computing 2018

% [liu21tsc, liu15] Compares supercapacitors versus lead-acid batteries. Develops strategies for assigning electrical load to these batteries.   

% BAAT: Towards Dynamically Managing Battery Aging in Green Datacenters. Liu et al. DSN 2015.

% [liu15dsn] Characterizes battery aging and distributes load across batteries. 

% Leveraging distributed UPS energy for managing solar energy powered data centers. Liu et al. IGCC 2014.
% RE-UPS: an adaptive distributed energy storage system for dynamically managing solar energy in green datacenters

% [liu16jsc, liu14igcc] Develops battery charge/discharge policies based on availability of time-of-day and availability of solar. Assumes solar is on-site. 

\textbf{Energy Storage.} Batteries ensure datacenter availability but can also modulate the datacenter's demands for grid power \cite{govindan12, govindan14}. For datacenters that use renewable energy, batteries can mitigate intermittent supplies of solar and wind \cite{liu2016jsc, liu14igcc}. Performance and efficiency vary with battery technology, motivating heterogeneous solutions \cite{liu21tsc, liu15}. Battery aging can be mitigated by managing charge-discharge cycles and demand for stored energy \cite{liu15dsn}. We quantify energy storage required for 24/7 carbon-free computing and, without loss of generality, consider lithium-ion batteries for their attractive downward cost trajectory and acceptable ten-plus year lifetimes under simulated usage.

Battery technologies will impact data center design and management. The price of lithium-ion batteries is falling significantly, declining by 80\% from 2015 to 2020~\cite{nrel20, doe19}. These batteries have been deployed at scale and, for example, can supply 28 MW for four hours. Such operational parameters align with hyperscale datacenters, which are provisioned for 20 to 40 MW. Four hours of battery operation could significantly reduce demand response requirements from job scheduling.

%For any given level of data center demand response, there exists an equivalent energy storage solution. When carbon-free energy is scarce, the data center can either defer computation or draw on energy storage. A prior case study has found that, from the grid’s perspective, a 20MW data center with 20\% demand flexibility is equivalent to a data center with 0.67MWh of storage when managing the grid’s voltage emergencies significantly, declining by 80\% from 2015 to 2020~\cite{liu14}. Although the study makes a number of optimistic assumptions about battery dis/charge speed and parameters may have evolved in the past few years, the study highlights the close relationship between energy storage and demand response.

In terms of the implementation strategy of batteries, datacenter operators could collaborate with utility providers to invest in batteries on the grid just as they do for wind and solar farms. Alternatively, they could deploy batteries on-site at the datacenter. Datacenter may wish to implement custom battery charge-discharge policies, which have previously been explored at much smaller scales for uninterruptible power supplies~\cite{govindan14,govindan12}. Whether these policies can be implemented in the form of contracts with grid operators is to be determined. 

Finally, there are environmental and health risks associated with extraction of lithium from either the sea or the soil and the disposal of batteries.
Spent LIBs contain toxic materials including heavy metals and flammable electrolytes, and therefore they need to be properly recycled and disposed in order not to cause contamination of the soil, water and air~\cite{RAJAEIFAR2022106144,zeng2015solving}. This is another aspect that needs consideration when making large-scale battery deployments.

\textbf{Carbon-Aware Scheduling.} Time-series analysis accurately forecasts renewable supplies and datacenter demands for energy. Forecasts permit optimizing schedules of flexible jobs in response to energy supply \cite{zhang21a}. Optimization objectives have accounted for electricity prices \cite{liu14}, carbon prices in cap-and-trade markets \cite{le2010}, the carbon-intensity of grid energy \cite{radovanovic2021carbon}, and service quality \cite{katsak2015}. Timely energy data is necessary for intelligent scheduling \cite{bashir2021enabling, wierman14}. We perform offline analyses to defer flexible computation and explore the design space for 24/7 carbon-free computing. A future implementation would benefit from prior schedulers.

\textbf{Other Considerations.}
Note that Carbon Explorer quantifies the net impact on carbon emissions, the key contributor to climate change, and leaves broader considerations in sustainability for future work. Our view of lithium-ion batteries quantifies carbon from manufacturing and recycling but neglects the impact of lithium extraction or other side effects. Similarly, our view on servers and power do not consider the impact of electronic waste and water usage for cooling. Although important, these broader environmental impacts are beyond the scope of our study. 

Moreover, Carbon Explorer emphasizes parameterized models because our understanding of carbon emissions in computing is still rapidly evolving. Operational emissions depend on the nature of renewable energy investments, the financial agreements for procuring this energy, and battery technologies. Embodied emissions depend on accurate accounting and transparent reporting throughout a massive supply chain. Carbon Explorer sets parameters based on the best publicly available data and these parameters can be tuned as better data becomes available.

\section{Conclusion}
This paper presents \textit{Carbon Explorer} --- a design space exploration tool to enable carbon-optimal investment strategies. Carbon Explorer determines carbon-optimal settings across the dimensions of \textit{investments on various renewable energy types}, \textit{the amount of energy storage}, and \textit{carbon-aware computation shifting} by considering geographically-dependent renewable energy availability characteristics and computation demand patterns at the data center scale.   
\CarbonExplorer demonstrates that, depending on graphical locations, carbon-optimal strategies vary
and that when embodied carbon
footprint is considered, 100\% 24/7 operational carbon-free computing may not always be carbon-optimal.
We hope \CarbonExplorer can guide future sustainability investments to achieve operational and embodied carbon footprint optimality.

\begin{acks}
We would like to thank Aditya Sundarrajan, Jiali Xing, Hsien-Hsin Sean Lee, Dimitrios Skarlatos, Max Balandat, Newsha Ardalani, Hugh Leather, Sal Candido for discussions and feedback on this work. We also thank Kim Hazelwood for supporting this work.
\end{acks}
%%%%%%% -- PAPER CONTENT ENDS -- %%%%%%%%

%%%%%%%%% -- BIB STYLE AND FILE -- %%%%%%%%
\bibliographystyle{ACM-Reference-Format}
\bibliography{refs}
%%%%%%%%%%%%%%%%%%%%%%%%%%%%%%%%%%%%

\end{document}